\documentclass[a4paper,10pt]{article}

\usepackage{subfig}
\usepackage{graphicx}
\usepackage{cite}
\usepackage{physics}
\usepackage{epsfig,amsmath,amssymb,verbatim,mathrsfs,array,layout,textcomp,amssymb,latexsym}

\newcommand{\beq}{\begin{eqnarray}}
\newcommand{\eeq}{\end{eqnarray}}
\newcommand{\be}{\begin{eqnarray}}
\newcommand{\ee}{\end{eqnarray}}

\def\beqa{\begin{eqnarray}}
\def\eeqa{\end{eqnarray}}
\def\bea{\begin{eqnarray}}
\def\eea{\end{eqnarray}}

\newcommand{\bv}{\left(\begin{array}{c}}
\newcommand{\bmtwo}{\left(\begin{array}{cc}}
\newcommand{\bmthree}{\left(\begin{array}{ccc}}
\newcommand{\emn}{\end{array}\right)}
\newcommand{\bmtwoc}{\left\{\begin{array}{cc}}
\newcommand{\bmthreec}{\left\{\begin{array}{ccc}}
\newcommand{\emnc}{\end{array}\right\}}
\newcommand{\ba}{\begin{array}}
\newcommand{\ea}{\end{array}}

\newcommand\cmt[1]{}
\def\lsim{\mathrel{\rlap{\lower4pt\hbox{\hskip1pt$\sim$}}
     \raise1pt\hbox{$<$}}}         
\def\gsim{\mathrel{\rlap{\lower4pt\hbox{\hskip1pt$\sim$}}
     \raise1pt\hbox{$>$}}}         

\newcommand{\leri}[1]{\left(#1 \right)}
\usepackage[section]{placeins}   

\usepackage{xcolor}

\usepackage[nolist]{acronym}
\begin{acronym}
	\acro{SM}{Standard Model}
	\acro{CPV}{CP violating}
	\acro{CPC}{CP conserving}
	\acro{EH}{Euler-Heisenberg}
	\acro{BSM}{beyond the Standard Model}
	\acro{COM}{center of mass}
	\acro{QED}{Quantum Electrodynamics}
	\acro{EFT}{Effective Field Theory}
	\acro{ALP}{axion-like particle}
	\acro{EOM}{Equation of Motion}
	\acro{RHS}{right hand side}
	\acro{LHS}{left hand side}
	\acro{FP}{Fabry-Perot}
	\acro{EM}{Electromagnetism}
	\acro{SRF}{superconducting radio frequency}
	\acro{SNR}{signal to noise ratio}
	\acro{EP}{Equivalence Principle}
	\acro{CL}{confidence level}
	\acro{DM}{dark matter}
	\acro{ULDM}{ultra-light dark matter}
	\acro{VEV}{vacuum expectation value}
	\acro{EWSB}{Electroweak Symmetry Breaking}
\end{acronym}

\begin{document}

\begin{titlepage}

\vskip1.5cm
\begin{center}
{\Large \bf Time Dependent CP-even and CP-odd Signatures of Scalar Ultra-light Dark Matter in Neutrino Oscillations} \\
\end{center}
\vskip0.2cm

\begin{center}
Marta Losada$^1$, Yosef Nir$^2$, Gilad Perez$^2$, Inbar Savoray$^2$ and Yogev Shpilman$^2$\\

\end{center}
\vskip 8pt

\begin{center}
{ \it $^1$New York University Abu Dhabi, PO Box 129188, Saadiyat Island, Abu Dhabi, United Arab Emirates\\
$^2$Department of Particle Physics and Astrophysics,\\
Weizmann Institute of Science, Rehovot 7610001, Israel} \vspace*{0.3cm}

{\tt  marta.losada@nyu.edu, yosef.nir,gilad.perez,inbar.savoray, yogev.shpilman@weizmann.ac.il}
\end{center}

\vglue 0.3truecm

\begin{abstract}
\noindent
Scalar \ac{ULDM} interacting with neutrinos can induce, under certain conditions, time-dependent modifications to neutrino oscillation probabilities. The limit in which the \ac{ULDM} perturbation can be treated as constant throughout the neutrino propagation time has been addressed by several previous works. We complement these by systematically analyzing the opposite limit -- accounting for the temporal-variations of the \ac{ULDM} potential by solving time-dependent Schr\"odinger equations. In particular, we study a novel two-generations-like \ac{CPV} signature unique to rapidly oscillating \ac{ULDM}. We derive the leading order, time-dependent, corrections to the oscillation probabilities, both for \ac{CPC} and \ac{CPV} couplings, and explain how they can be measured in current and future experiments. 
\end{abstract}

\end{titlepage}

\acresetall
\section{Introduction}
Scalar \ac{ULDM} with mass $m_\phi\lsim{\rm eV}$, may leave signatures in neutrino oscillation experiments, that cannot be interpreted within the \ac{SM} supplemented by neutrino masses and lepton mixing. The scalar can be treated as a classical bosonic field that oscillates with time,
\beq
\label{eq:scalarOscillations}\phi=\phi_0\sin(m_\phi(t-t_0))\,,
\eeq
with an arbitrary initial time $t_0$ and amplitude
\beq
\phi_0=\frac{\sqrt{2\rho_\phi^\oplus}}{m_\phi}\sim2\ {\rm GeV}\ \left(\frac{10^{-12}\ {\rm eV}}{m_\phi}\right),
\eeq
where $\rho_\phi^\oplus$ is the local \ac{ULDM} density. Consider the effective mass and $\phi$-Yukawa terms for the neutrinos, arising from dimension-five and dimension-six terms in the Lagrangian, respectively. In the basis in which the \ac{ULDM}-independent mass matrix is diagonal, they read
\beq \label{eq: m+y}
{\cal L}_{m_\nu}=m_i \nu_i^T\nu_i + \hat y_{ij}\phi\nu_i^T\nu_j+{\rm h.c}\,.
\eeq
Treating $\phi$ as a classical field, it modifies the neutrino mass matrix as
\beq \label{eq: mass matrix}
(\hat m_\nu)_{ij}=m_i\delta_{ij}+\hat y_{ij}\phi\,,
\eeq
thus inducing a time-dependent component to the neutrino propagation Hamiltonian.

Some of the implications of the \ac{ULDM} contribution to the neutrino mass matrix have been studied in previous works~\cite{Krnjaic:2017zlz,Brdar:2017kbt,Capozzi:2018bps,Berlin:2016woy,Dev:2020kgz,Losada:2021bxx,Chun:2021ief,Huang:2021kam}. Most previous analyses focused on \ac{ULDM} candidates with typical oscillation times, $\tau_\phi\equiv 2\pi/m_\phi$, much larger than the neutrino propagation time, $\tau_d$, but smaller than the total run-time of the experiment, $\tau_e$~\cite{Berlin:2016woy,Krnjaic:2017zlz,Dev:2020kgz,Losada:2021bxx}. In this limit, the effective neutrino masses and mixing do not change during the propagation from the source to the detector, but do change between neutrino events over the duration of the experiment. Depending on the value of $m_\phi$, the experiment time resolution $\tau_r$, and the way one analyses the data, the \ac{ULDM} effect can be either time-resolved, or time-averaged. In the former case, the temporal modulation could be directly observed in some spectral analysis of the data~\cite{Dev:2020kgz,Losada:2021bxx}. In the latter case, the oscillatory contributions are averaged out, and one must use indirect methods for recovering evidence of the time dependence, e.g. via the distortion of the dynamical range of the neutrino oscillation parameters or of relations between oscillation amplitudes at different $L/E$ frequencies~\cite{Krnjaic:2017zlz,Chun:2021ief,Dev:2020kgz,Losada:2021bxx} ($L$ is the source to detector distance and $E$ is the neutrino energy).

However, the experimental signatures of heavier \ac{ULDM}, with $\tau_\phi\lesssim \tau_d$, require a dedicated analysis. In the limit of rapid oscillations, one must solve time-dependent equations of motion to correctly account for the \ac{ULDM} effects on neutrino oscillation probabilities, and the picture of slowly varying oscillation parameters is no longer valid~\cite{Dev:2020kgz,Losada:2022uvr}. In addition, thus far, the majority of previous works discussing \ac{ULDM}-modified neutrino oscillations have only considered \ac{CPC} couplings. \ac{CPV} effects in neutrino oscillations induced by \ac{ULDM}-neutrino interactions have been analyzed by authors of this manuscript in Ref.~\cite{Losada:2021bxx}, for the case of slow oscillations. Furthermore, past works have either studied specific flavor structures for the \ac{ULDM} couplings~\cite{Brdar:2017kbt,Chun:2021ief,Losada:2021bxx}, or studied the ``model-agnostic" temporal-modulations of the mixing angles, mass differences and the \ac{CPV} phase of the PMNS matrix~\cite{Krnjaic:2017zlz,Dev:2020kgz,Losada:2021bxx}. To the best of our knowledge, the relations between the \ac{ULDM} couplings and these variations of oscillation parameters have not been explicitly discussed, and could be non trivial (both mathematically and phenomenologically). Therefore, translating the conclusions of the existing analyses to generic \ac{ULDM} models could be challenging.

In this work, we fill in those gaps, and provide a more complete description of the phenomenology of \ac{ULDM}-neutrino couplings. We are interested in mapping those couplings onto their signatures in neutrino oscillation experiments, for a wide range of \ac{ULDM} masses, covering both the slow oscillations and the rapid oscillations limits. We present a generic prescription for estimating the first-order, time-dependent, modifications to neutrino oscillations from \ac{ULDM} couplings to neutrinos, and the resulting bounds on them from current and future experiments. While we study both \ac{CPC} and \ac{CPV} phenomena, we give special attention to the implications of the \ac{CPV} \ac{ULDM} couplings. In particular, we describe new \ac{CPV} effects generated by them, that can be observable even in two-generations neutrino systems, and cannot be addressed by analyses assuming a constant neutrino Hamiltonian through propagation.

The paper is organized as follows. In Sec.~\ref{sec:2_nu_CPV}, we describe the new \ac{CPV} effect in the two neutrino generations picture - which we refer to as the ``2-$\nu$ \ac{CPV}". In Sec.~\ref{sec:analytic_expressions}, we derive the full analytical expressions for the leading order modifications to neutrino oscillation probabilities induced by the \ac{ULDM}, both for \ac{CPC} and \ac{CPV} phenomena. In Sec.~\ref{sec: Rayleigh}, we describe the method for detecting such time-dependent modifications in neutrino oscillations data. In Sec.~\ref{sec:bounds}, we show the current bounds, and the future-projected sensitivities, for the \ac{ULDM} parameter space. We conclude in Sec.~\ref{sec:conclusions}.

\section{$2-\nu$ CP violation}\label{sec:2_nu_CPV}

We are interested in the possibility of observing \ac{CPV} in neutrino oscillations induced by the time-dependent \ac{VEV} of $\phi$. In the usual two-generations picture, one can show that by field redefinitions, the mass matrix $\hat{m}_\nu$ can always be brought to a form in which it has no physical phases affecting neutrino oscillations~(see e.g.~\cite{BILENKY1980495,Giunti:2010ec}), and thus \ac{CPV} cannot be observed in these processes. This can be seen by the simple following argument. In the two-neutrino picture, the \ac{EOM} is given by
\begin{align}
\label{eq:nuEOM}
i\partial_{t}\left(\begin{matrix}\nu_{\alpha}\\
\nu_{\beta}
\end{matrix}\right) & ={\cal H}\left(t\right)\left(\begin{matrix}\nu_{\alpha}\\
\nu_{\beta}
\end{matrix}\right)=\left(\begin{matrix}{\cal H}_{11} & {\cal H}_{12}e^{i\varphi}\\
{\cal H}_{12}e^{-i\varphi} & {\cal H}_{22}
\end{matrix}\right)\left(\begin{matrix}\nu_{\alpha}\\
\nu_{\beta}
\end{matrix}\right)\,.
\end{align}
Since the Hamiltonian ${\cal H}$ is Hermitian, ${\cal H}_{ij}$ and $\varphi$ above must be real. Let $\nu_i(t)$ be the time-evolved state, according to Eq.~\eqref{eq:nuEOM}, which was $\nu_i$ at $t=0$. If $\nu_{\alpha,\beta}$ are interaction eigenstates, then the initial conditions at $t=0$ are zero for one of them, and one for the other, and the probability for detecting the states $\nu_{\alpha,\beta}$ at some later time $t$ is given by solving the \ac{EOM} and calculating 
\begin{align}
\label{eq:Pnunu}
P_{\nu_i\rightarrow\nu_f}=|\langle \nu_{f}|\nu_i(t)\rangle|^2\,.
\end{align} 
While ${\cal H}$ carries a \ac{CPV} Majorana phase $\varphi$, it can be absorbed in the redefinition of the field $\nu'_\beta=e^{i\varphi}\nu_\beta$, which makes the \ac{EOM} real
\begin{align}
i\partial_{t}\left(\begin{matrix}\nu_{\alpha}\\
\nu'_{\beta}
\end{matrix}\right) & ={\cal H}'\left(t\right)\left(\begin{matrix}\nu_{\alpha}\\
\nu'_{\beta}
\end{matrix}\right)=\left(\begin{matrix}{\cal H}_{11} & {\cal H}_{12}\\
{\cal H}_{12} & {\cal H}_{22}
\end{matrix}\right)\left(\begin{matrix}\nu_{\alpha}\\
\nu'_{\beta}
\end{matrix}\right)\,.
\end{align}
In the context of neutrino oscillations, this field redefinition has no physical significance, since $\nu'_\beta$ is still a flavor eigenstate, and the oscillation probabilities for any $\nu'$ and $\nu$ states that are related by arbitrary phases $\nu'_{\alpha,\beta}=e^{i\varphi_{\alpha,\beta}} \nu_{\alpha,\beta}$ are the same, since
\begin{align}
\label{eq:Pnu'nu}
P_{\nu'_i\rightarrow\nu'_f}=|\langle \nu'_{f}|\nu'_i(t)\rangle|^2=|\langle \nu_{f}|e^{i(\varphi_i-\varphi_f)}|\nu_i(t)\rangle|^2=P_{\nu_i\rightarrow\nu_f}\,.
\end{align}
For anti-neutrinos, which are the CP conjugates of $\nu$, i.e. ${\bar \nu}=i\sigma_2\nu^*$, the \ac{EOM} is given by
\begin{align}
i\partial_{t}\left(\begin{matrix}\overline{\nu}_{\alpha}\\
\overline{\nu}_{\beta}
\end{matrix}\right) & ={\cal H}^{*}\left(t\right)\left(\begin{matrix}\overline{\nu}_{\alpha}\\
\overline{\nu}_{\beta}
\end{matrix}\right)\,,
\end{align}
and thus we may again redefine our fields as $\overline{\nu'}_\beta=e^{-i\varphi}\overline{\nu}_\beta$ and obtain
\begin{align}
i\partial_{t}\left(\begin{matrix}\overline{\nu}_{\alpha}\\
\overline{\nu'}_{\beta}
\end{matrix}\right) & ={\cal H}'\left(t\right)\left(\begin{matrix}\overline{\nu}_{\alpha}\\
\overline{\nu'}_{\beta}
\end{matrix}\right).
\end{align}
We then learn that, since ${\cal H'}$ does not depend on $\varphi$, anti-neutrinos propagate like neutrinos.

In the presence of the oscillating \ac{ULDM} background, $\phi(t)$, the \ac{CPV} phase, $\varphi$, could become time-dependent. Then, the field redefinition $\nu'_\beta=e^{i\varphi(t)}\nu_\beta$ yields
\begin{align}
\label{eq:EOMCPVnu}
i\partial_{t}\left(\begin{matrix}\nu_{\alpha}\\
\nu'_{\beta}
\end{matrix}\right) & =\left(\begin{matrix}{\cal H}_{11} & {\cal H}_{12}\\
{\cal H}_{12} & {\cal H}_{22}-\dot{\varphi}
\end{matrix}\right)\left(\begin{matrix}\nu_{\alpha}\\
\nu'_{\beta}
\end{matrix}\right)\,,
\end{align}
for neutrinos, and
\begin{align}
\label{eq:EOMCPVnubar}
i\partial_{t}\left(\begin{matrix}\overline{\nu}_{\alpha}\\
\overline{\nu'}_{\beta}
\end{matrix}\right) & =\left(\begin{matrix}{\cal H}_{11} & {\cal H}_{12}\\
{\cal H}_{12} & {\cal H}_{22}+\dot{\varphi}
\end{matrix}\right)\left(\begin{matrix}\overline{\nu}_{\alpha}\\
\overline{\nu'}_{\beta}
\end{matrix}\right)\,,
\end{align}
for anti-neutrinos. Note that the oscillation probabilities for $\nu$ and $\nu'$ are still identical, since equation \eqref{eq:Pnu'nu} still holds even when promoting $\varphi$ to be time-dependent. Thus, neutrinos and anti-neutrinos would evolve differently in time, as a result of the time-dependent Majorana phase. 

While the instantaneous phase of the fields is not physically observable, its derivative -- which can be thought of as an instantaneous frequency -- shifts the energy of the system, and is thus observable. Indeed, the Hamiltonians in Eq.~\eqref{eq:EOMCPVnu} and Eq.~\eqref{eq:EOMCPVnubar} are very similar to those resulting from charged current matter effects, identifying $-\dot{\varphi}$ with the matter potential $V_C$ \cite{PhysRevD.17.2369}. Note that in our case $-\dot{\varphi}$ is a periodic function of time, which will ultimately result in a time periodic CP asymmetry.

For concreteness, let us explicitly examine this effect in a two generations system. Consider the following Yukawa matrix $\hat{y}$ describing the \ac{ULDM} coupling to the unperturbed mass eigenstates in two generations,
   	\begin{align}
   	\hat{y} & =iy_I\left(\begin{matrix}0 & 1\\
   	1 & 0
   	\end{matrix}\right),
   	\end{align}
   	with real $y_I$. Given the Yukawa interaction in Eq.~\eqref{eq: m+y} is Majorana-like, the most general form of $\hat{y}$ is
   	$\hat y= a_\mu\sigma_\mu$, with $\mu=0,1,3\,,$ $\sigma_0=\mathbf{1}_2$, $\sigma_{1,3}$ are the corresponding symmetric Pauli matrices, and the $a_\mu$'s are complex numbers.
	Note that within the two generation framework, at any given time the unperturbed instantaneous PMNS matrix can be brought to be real, and so it is straightforward to show that the form of $\hat{y}$ written above uniquely leads to two-generation \ac{CPV}.
	In the interaction basis, in the ultra-relativistic limit, the Hamiltonian can be written as
   	\begin{align}
   	{\cal H} &=\frac{1}{2E}\hat m_\nu^\dagger\hat m_\nu 
   	 \sim\frac{\Delta m^{2}}{4E}\left(\begin{matrix}-\cos2\theta & \sin2\theta\\
   			\sin2\theta & \cos2\theta
   			\end{matrix}\right)
   	+\frac{\phi_{0}}{2E}\left(\begin{matrix}0 & -i y_{I} \Delta m\\
   			iy_{I} \Delta m  & 0
   			\end{matrix}\right)\sin\left(m_{\phi}\leri{t-t_0}\right)+{\cal O}\left(\frac{\phi_{0}^{2}}{E}\right),
   	\end{align}
   	where $\theta$ is the mixing angle between the interaction eigenstates in the unperturbed mass eigenstates, and $\Delta m=m_{2}-m_{1}$, $\Delta m^{2}=m_{2}^{2}-m_{1}^{2}$. To leading order in the \ac{ULDM} perturbation, we identify 
   \begin{align}
   \dot{\varphi}\approx-\frac{2 m_\phi\phi_0 y_I}{ \leri{m_1+m_2}\sin{2\theta}}\cos\leri{m_\phi\leri{t-t_0}}\,.
   \end{align}
Following the prescription presented in the next section, the leading order difference between the survival probability of neutrinos $P_{\alpha\alpha}$ and the survival probability of anti-neutrinos $P_{\bar{\alpha}\bar{\alpha}}$ is given by 
   \begin{align}
   \Delta P^{(1)}_{\alpha\alpha}  & =\frac{2\phi_0 \sin{4\theta} y_{I}\leri{m_1-m_2}}{E\leri{\leri{\frac{\Delta m^{2}}{2E}}^2-m_\phi^2}}\times\nonumber\\
   \times\Big[&\cos\leri{m_\phi \leri{\leri{t-L-t_0}+\frac{L}{2}}}\leri{\frac{\Delta m^{2}}{4E}\sin{\frac{\Delta m^{2}L}{2E}}\sin{\frac{m_\phi L}{2}}-m_\phi \sin^2{\frac{\Delta m^{2}L}{4 E}}\cos{\frac{m_\phi L}{2}}}\Big ]\,.\label{eq:2nuCPV}
   \end{align}
In the following sections, we discuss the implications of this ``2-$\nu$ \ac{CPV}" effect, and how it can be searched for in neutrino oscillations experiments.
   
\section{Analytic approximation for small perturbations}\label{sec:analytic_expressions}

\subsection{Generic Prescription}
The neutrino Schr\"odinger \acp{EOM} can be written as
\begin{align}
i\partial_t \nu\leri{t}=\frac{1}{2E}{\hat m}^\dagger \hat{m}= \frac{1}{2E}\leri{m^\dagger m+\leri{m^\dagger y +y^\dagger m}\phi\leri{t} +y^\dagger y \phi^2\leri{t}}\,.~\label{eq:tdep_EOM}
\end{align}

The solutions of systems with time dependent Hamiltonians $\mathcal{H}(t)=\mathcal{H}_0+V(t)$ can be found using the Dyson series. A neutrino state at any final time point $t_f$, is evolved from an initial neutrino state at time $t_i$ as $\ket{\nu\leri{t_f}} =\mathcal{U}(t_f,t_i)\ket{\nu\leri{t_i}}$ with
\begin{align}
\mathcal{U}(t_f,t_i)=\sum_{n=0}^{\infty} \mathcal{U}^{(n)}(t_f,t_i)\equiv\sum_{n=0}^{\infty}\left(-i\right)^{n}\int\limits _{t_f\geq t_{1}...\geq t_{n}\geq t_i}dt_{1}...dt_{n}e^{-i{\cal H}_{0}\left(t-t_{1}\right)}V(t_1)e^{-i{\cal H}_{0}\left(t_{1}-t_{2}\right)}...V(t_n)e^{-i{\cal H}_{0}\leri{t_{n}-t_i}}\,.
\end{align}
The probability for a neutrino produced in a state $\nu_\alpha$ at time $t_i$ to be measured as neutrino state $\nu_\beta$ at a later time $t_f$ is 
\begin{align}
P_{\alpha\beta}\left(t_f\right) & =\left|\left\langle \nu_{\beta}| \mathcal{U}\leri{t_f,t_i}|\nu_{\alpha}\right\rangle \right|^{2}\equiv \left| \mathcal{U}_{\beta \alpha}\leri{t_f,t_i}\right|^{2}\,.
\end{align}

In our case, we identify ${\cal H}_{0}$ with the \ac{ULDM}-independent contribution to Eq.~\eqref{eq:tdep_EOM}, and denote $V(t)=V^{(1)}(t)+V^{(2)}\leri{t}$ with
\begin{align}
V^{(1)}(t)&\equiv \frac{\phi_0}{2E}\tilde{m}\sin(m_\phi(t-t_0))\,,\label{eq:V1}\\
V^{(2)}(t)&\equiv \frac{\phi^2_0}{2E} y^\dagger y\sin^2(m_\phi(t-t_0))\,,\label{eq:V2}
\end{align}
where we define $\tilde{m}\equiv m^\dagger_\nu\hat{y}+{\hat{y}}^\dagger m_\nu$\,.

Then, the \ac{ULDM}-independent probabilities $P^{(0)}_{\alpha\beta}$ are given by 
\begin{align}\label{eq:P0}
P^{(0)}_{\alpha\beta}=\left| \mathcal{U}^{(0)}_{\beta \alpha}\leri{t_f,t_i}\right|^{2} \,,
\end{align}
where 
\begin{align}
\mathcal{U}^{(0)} \leri{t_f,t_i}& =e^{-i{\cal H}_{0}\leri{t_f-t_i}}\,.
\end{align}
As expected $\mathcal{U}^{(0)}\leri{t_f,t_i}=\mathcal{U}^{(0)}\leri{t_f-t_i}$. From this point onwards, we assume $t_f-t_i$ corresponds to the propagation time over a distance $L$, i.e $t_f-t_i\approx L$. Thus $\mathcal{U}^{(0)}$ and $P^{(0)}$ are functions of $L$ alone, and do not vary with time. For the sake of brevity, moving forward we keep the $L$ dependence implicit, and only explicitly note time dependencies. In addition, in the following we assume $\alpha$ and $\beta$ are interaction eigenstates.

The leading order perturbation to the transition/survival probabilities is given by (no summation is implied on $\alpha,\beta$)
\begin{align}\label{eq:P1}
P^{(1)}_{\alpha\beta}(t)& =2\text{Re}\leri{\mathcal{U}_{\beta\alpha}^{(0)}{\mathcal{U}_{\beta\alpha}^{(1)}\leri{t}}^*}\,,
\end{align} 
where 
\begin{align}
\label{eq:U1Dyson}
\mathcal{U}^{(1)}\leri{t} & =-i\int\limits _{t-L}^{t}dt_{1}e^{-i{\cal H}_{0}\left(t-t_{1}\right)}V^{(1)}\leri{t_1}e^{-i{\cal H}_{0}\leri{t_1-t+L}}=-i\int\limits _{0}^{L}dt_{1}e^{-i{\cal H}_{0}\left(L-t_{1}\right)}V^{(1)}\leri{t_1+t-L}e^{-i{\cal H}_{0}t_1}\,.
\end{align}
Since $V^{(1)}$ given in Eq.~\eqref{eq:V1} is sinusoidal, the integral is easily solved analytically, and we obtain 
\begin{align}
P^{(1)}_{\alpha\beta}\leri{t}& =\frac{2 \phi_0}{ E}\text{Im}\leri{U_{\beta k}U^*_{\alpha k}U^*_{\beta i}U_{\alpha j} {\tilde{m}_{ij}}^*e^{i\frac{\leri{\Delta m^{2}_{jk}+\Delta m^{2}_{ik}}L}{4E}}{\kappa^*\leri{t}_{ij}}}\,,\label{eq:P1alphabetaGeneric}
\end{align}
where the $i,j,k$ indices correspond to the eigenstates of the \ac{ULDM}-independent Hamiltonian $\mathcal{H}_0$, and $U$ is the \ac{ULDM}-independent PMNS matrix, namely - the transition matrix between the interaction eigenstates and the eigenstates of $\mathcal{H}_0$. Specifying the form of the \ac{ULDM} couplings $\hat{y}_{ij}$ in the basis in which $\mathcal{H}_0$ is diagonal, 
\begin{align}
\tilde{m}_{ij}=\leri{m^\dagger_\nu\hat{y}+{\hat{y}}^\dagger m_\nu}_{ij}=m_i\hat{y}_{ij}+{\hat{y}_{ij}}^* m_j=\text{Re}\leri{\hat{y}_{ij}}\leri{m_i+m_j}+i\text{Im}\leri{\hat{y}_{ij}}\leri{m_i-m_j}\,.
\end{align}
Lastly, $\kappa_{ij}\leri{t}$ is complex, with phases that are only dynamical (\ac{CPC}). Its real and imaginary parts are given by
\begin{align}
\label{eq:Rek}\text{Re}\leri{\kappa_{ij}}&=\frac{\sin\leri{m_\phi \leri{t-\frac{L}{2}-t_0}}}{\leri{\frac{\Delta m^{2}_{ij}}{2E}}^2-m_\phi^2}\leri{ m_\phi\cos{\frac{\Delta m^{2}_{ij}L}{4 E}}\sin{\frac{m_\phi L}{2}}-\frac{\Delta m^{2}_{ij}}{2E}\sin{\frac{\Delta m^{2}_{ij}L}{4 E}}\cos{\frac{m_\phi L}{2}}}\,,\\
\label{eq:Imk}\text{Im}\leri{\kappa_{ij}}&=\frac{\cos\leri{m_\phi \leri{t-\frac{L}{2}-t_0}}}{\leri{\frac{\Delta m^{2}_{ij}}{2E}}^2-m_\phi^2}\leri{\frac{\Delta m^{2}_{ij}}{2E}\cos{\frac{\Delta m^{2}_{ij}L}{4 E}}\sin{\frac{m_\phi L}{2}}-m_\phi \sin{\frac{\Delta m^{2}_{ij}L}{4 E}}\cos{\frac{m_\phi L}{2}}}\,.
\end{align}

From the temporal dependence of $\kappa$, and thus also of the neutrino probabilities, we see that simply adding up data points gathered over times larger than $m_\phi^{-1}$ would suppress the effect due to averaging over different neutrino arrival times $t$. We therefore require time-stamping the events with a minimal temporal resolution corresponding to $\sim 1/m_\phi$. While most previous analyses assumed the \ac{ULDM} field to be constant during the propagation time $L$, we are interested in extending these results into the $m_\phi L\gtrsim 1$ region. For long baseline neutrino experiments, this is equivalent to
\begin{equation}
m_{\phi}\gtrsim10^{-12}\text{ eV},
\end{equation}
or
\begin{equation}
\tau_{\phi}\equiv\frac{2\pi}{m_{\phi}}\lesssim1\text{ ms}.
\end{equation}
This time scale is much shorter than the typical time between events in any neutrino experiment. In order to study how to probe these time variations, we will use the unbinned Rayleigh periodogram, as will be explained in Sec.~\ref{sec: Rayleigh}.

Note that in the limit $m_\phi L\rightarrow 0\,, 2Em_\phi/ \Delta m_{ij}^2\rightarrow 0\,$, we have $\text{Im}\leri{\kappa_{ij}}/\text{Re}\leri{\kappa_{ij}}\rightarrow 0$ ($\text{Re}\leri{\kappa_{ij}}$ remains finite as $V\leri{t}\propto \sin{m_\phi \leri{t-t_0}}$). Therefore, previous analyses that assumed the \ac{ULDM} potential to be constant within the propagation time, have effectively only taken into account effects that are associated with the real part of $\kappa$, which in that limit reduces to
\begin{align}
\label{eq:reKappaConstULDM}
\text{Re}\leri{\kappa_{ij}}&\approx-\frac{\sin\leri{m_\phi \leri{t-t_0}}}{\frac{\Delta m^{2}_{ij}}{2E}}\sin{\frac{\Delta m^{2}_{ij}L}{4 E}}\,.
\end{align}
In this limit, for \ac{ULDM} couplings that are off-diagonal in the unperturbed mass basis, the result can be interpreted as following from the first order correction to the diagonalizing matrix of the neutrino Hamiltonian in the interaction basis. Namely, 
\begin{align}
\label{eq:UtildeConst}
U_{\alpha k}\rightarrow \tilde{U}_{\alpha k} \leri{t} =  \leri{U T^\dagger\leri{t}}_{\alpha k}\,,
\end{align}
where 
\begin{align}
\label{eq:TConst}
T\leri{t}_{ij}=\delta_{ij}+\leri{1-\delta_{ij}}\frac{\phi_0\tilde{m}_{ij}}{\Delta m^2_{ij}}\sin\leri{m_\phi\leri{t-t_0}}\,.
\end{align}
Note that in this limit, the new diagonalizing matrix $\tilde{U}$ is treated as constant within propagation, but changes its value for each measurement at time $t$. 

To understand what happens when going beyond the constant potential limit, let us rewrite the \ac{EOM} in the interaction basis as
\begin{align}
i\partial_t \nu^I\leri{t} =U T^\dagger(t)\mathcal{H}_{D}T(t)U^{\dagger}\nu^I\leri{t}\approx U T^\dagger(t)\mathcal{H}_{0}T(t)U^{\dagger}\nu^I\leri{t}\,,
\end{align}
where $\nu^I$ is given in the interaction basis, and $\mathcal{H}_D$ is the instantaneously diagonal Hamiltonian, which to linear order in the \ac{ULDM} perturbation is just $\mathcal{H}_0$\,. To solve the \ac{EOM}, one may define
\begin{align}
\tilde{\nu}\leri{t}= {\tilde{U}\leri{t}}^{\dagger}\nu^I\leri{t} \,,
\end{align}
and obtain
\begin{align}
i\partial_t\tilde{\nu}\leri{t}= \leri{\mathcal{H}_0 -i{\tilde{U}\leri{t}}^{\dagger}\partial_t \tilde{U}\leri{t}}\tilde{\nu}\leri{t}=\leri{\mathcal{H}_0 -iT\leri{t}\partial_t {T\leri{t}}^{\dagger}}\tilde{\nu}\leri{t}\,.
\end{align}
As the \ac{EOM} is not diagonal due to the contributions coming from the time-derivative of the Hamiltonian, one should in principle diagonlize the new effective Hamiltonian by another time-dependent matrix, the derivative of which will again contribute to the Hamiltonian which should again be diagonalized, and so on. Keeping only the linear terms in the \ac{ULDM} perturbation, this procedure will yield an infinite series in $\partial^{(n)}_t V^{(1)}_{ij} /\leri{\Delta m^2_{ij}}^n$, where $i\neq j$ and $V^{(1)}$ is defined in Eq.~\eqref{eq:V1}. The result of this series is the effective diagonalizing matrix
\begin{align}
\label{eq:UtildeTot}
\tilde{U}\leri{t}= U\tilde{T}^\dagger\leri{t}\,,
\end{align} 
with
\begin{align}
\label{eq:TtildeTot}
\tilde{T}_{ij}\leri{t}=\delta_{ij}+\leri{1-\delta_{ij}}\frac{\phi_0\tilde{m}_{ij}}{\Delta m^2_{ij}}\sum_{n=0}^\infty i^{n}\leri{\frac{2E}{\Delta m_{ij}^2}}^n\partial^{(n)}_t\sin\leri{m_\phi\leri{t-t_0}}\,.
\end{align}
Since our potential is harmonic, we can easily sum all the contributions up to infinity, and obtain
\begin{align}
\tilde{T}_{ij}\leri{t}=\delta_{ij}+\leri{1-\delta_{ij}}\frac{\phi_0\tilde{m}_{ij}}{\Delta m^2_{ij}}\frac{1}{1-\leri{\frac{2E m_\phi}{\Delta m_{ij}^2}}^2}\leri{\sin\leri{m_\phi\leri{t-t_0}}+i\frac{2E m_\phi}{\Delta m_{ij}^2}\cos\leri{m_\phi\leri{t-t_0}}}\,.
\end{align} 
Therefore, the neutrino flavor probabilities are given by taking the zeroth and first order terms in $\phi_0\tilde m_{ij}/\Delta m^2_{ij}$ of
\begin{align}
P_{\alpha \beta}\leri{t}\approx  \tilde{U}_{\beta k}\leri{t} \tilde{U}^*_{\alpha k}\leri{t-L} \tilde{U}^*_{\beta i }\leri{t} \tilde{U}_{\alpha i}\leri{t-L}e^{-i \frac{\Delta m^2_{ki}L}{2E} } \,.
\end{align}

In the next subsections, we describe the full \ac{ULDM} effects in three generations, first considering \ac{CPC} observables in subsection~\ref{subsec:CPC}, and then \ac{CPV} phenomena in subsection~\ref{subsec:CPV}. Note, however, that the expressions above are completely generic for any number of generations, provided the indices are appropriately summed. The probabilities for the two generations case are presented in Appendix~\ref{appendix:2nuP}.

\subsection{CPC effects}
\label{subsec:CPC}
In the standard case with three neutrino generations, the neutrino probabilities in three generations are given by 
\begin{align}
P^{(0)}_{\alpha\beta}&=U_{\beta k}U^*_{\alpha k}U^*_{\beta i}U_{\alpha i}e^{-i\frac{\Delta m^2_{ki}L}{2E}}\,.
\end{align}
The \ac{CPC} part of this probability is given by averaging over neutrinos and anti-neutrinos 
\begin{align}
\Sigma P^{(0)}_{\alpha\beta}&\equiv \frac12\leri{P^{(0)}_{\alpha\beta}+P^{(0)}_{\bar{\alpha}\bar{\beta}}}=\text{Re}\leri{U_{\beta k}U^*_{\alpha k}U^*_{\beta i}U_{\alpha i}}\cos\leri{\frac{\Delta m^2_{ki}L}{2E}}\,.
\end{align}
We now discuss the first order corrections from \ac{ULDM} to \ac{CPC} neutrino probabilities. From Eq.~\eqref{eq:P1alphabetaGeneric}, we find \ac{CPC} contributions to be associated with the real parts of the complex matrices $U$ and $\hat{y}$. Let us denote these contributions by $\Sigma P^{(1)}_{\alpha\beta}$:
\begin{align}
\label{eq:SigmaP1}
\Sigma P^{(1)}_{\alpha\beta}\equiv \frac12\leri{P^{(1)}_{\alpha\beta}+P^{(1)}_{\bar{\alpha}\bar{\beta}}}&=\frac{2 \phi_0}{ E}\text{Re}\leri{U_{\beta k}U^*_{\alpha k}U^*_{\beta i}U_{\alpha j} {\tilde{m}_{ij}}^*}\text{Im}\leri{e^{i\frac{\leri{\Delta m^{2}_{jk}+\Delta m^{2}_{ik}}L}{4E}}{\kappa^*\leri{t}_{ij}}}\,.
\end{align}

In three generations, for diagonal couplings $\hat{y}_{ii}$, both $\tilde{m}_{ii}$ and $\kappa_{ii}$ are real, and thus for a specific $i$, specific $ j\neq k\neq i$ and any $\alpha,\beta$
\begin{align}
\label{eq:CPCdiagonal}
\Sigma P^{(1)}_{\alpha\beta}\leri{t}
&=-\sin\leri{m_\phi \leri{t-\frac{L}{2}-t_0}}\sin\leri{\frac{m_\phi L}{2}}\frac{2\tilde{m}_{ii} \phi_0}{ E m_\phi}\times\nonumber\\
&\times\leri{\text{Re}\leri{U_{\beta k}U^*_{\alpha k}U^*_{\beta i}U_{\alpha i}}\sin{\leri{\frac{\Delta m^{2}_{ik}L}{2E}}}+\text{Re}\leri{U_{\beta j}U^*_{\alpha j}U^*_{\beta i}U_{\alpha i}}\sin{\leri{\frac{\Delta m^{2}_{ij}L}{2E}}}}\,.
\end{align}
This result is nothing but an effective, time-dependent, shift of the neutrino mass differences $\Delta m^2_{ij}$
\begin{align}
\label{eq:diagonalDeltam}
\Delta m^2_{ij}\rightarrow \Delta m^2_{ij}+\frac{2m_i\phi_0\hat{y}_{ii}}{L}\int^t_{t-L} \sin\leri{m_\phi\leri{t_1-t_0}}dt_1\,.
\end{align}
Effects of this sort have already been quite thoroughly discussed in previous works such as~\cite{Krnjaic:2017zlz,Brdar:2017kbt,Chun:2021ief,Losada:2021bxx}.

For off-diagonal couplings $\hat{y}_{ij}$, with $i\neq j\neq k$
transition probabilities ($\alpha\neq\beta$) are modified by
\begin{align}\label{eq: CPC transition P1}
\Sigma P^{(1)}_{\alpha\beta} =\frac{4 \phi_0}{ E}\text{Im}\leri{\kappa^*\leri{t}_{ij}}\,&\text{Re}\leri{U_{\beta k}U^*_{\alpha k}\leri{U^*_{\beta i}U_{\alpha j}{\tilde{m}_{ij}}^*-U^*_{\beta j}U_{\alpha i}{\tilde{m}_{ij}} }}\sin{\frac{\Delta m^{2}_{ik}L}{4E}}\sin{\frac{\Delta m^{2}_{kj}L}{4E}}+\nonumber\\
\frac{4 \phi_0}{ E}\text{Re}\leri{\kappa^*\leri{t}_{ij}}\Bigg[&\text{Re}\leri{U_{\beta j}U^*_{\alpha j}\leri{U^*_{\beta i}U_{\alpha j}{\tilde{m}_{ij}}^*+U^*_{\beta j}U_{\alpha i}{\tilde{m}_{ij}} }}\sin{\frac{\Delta m^{2}_{kj}L}{4E}}\cos{\frac{\Delta m^{2}_{ki}L}{4E}}\nonumber\\
+&\,\text{Re}\leri{U_{\beta i}U^*_{\alpha i}\leri{U^*_{\beta i}U_{\alpha j}{\tilde{m}_{ij}}^*+U^*_{\beta j}U_{\alpha i}{\tilde{m}_{ij}} }}\sin{\frac{\Delta m^{2}_{ki}L}{4E}}\cos{\frac{\Delta m^{2}_{kj}L}{4E}}\Bigg]
\,,
\end{align}
and survival probabilities ($\alpha=\beta$) by
\begin{align}\label{eq: CPC Survival P1}
\Sigma P^{(1)}_{\alpha\alpha}& =
\frac{4 \phi_0}{ E}\text{Re}\leri{\kappa\leri{t}_{ij}}\text{Re}\leri{U^*_{\alpha i}U_{\alpha j}{\tilde{m}_{ij}}^* }\times\nonumber\\
&\times\Bigg[\leri{2|U_{\alpha j}|^2-1} \sin{\frac{\Delta m^{2}_{kj}L}{4E}}\cos{\frac{\Delta m^{2}_{ki}L}{4E}}
+\leri{2|U_{\alpha i}|^2-1} \sin{\frac{\Delta m^{2}_{ki}L}{4E}}\cos{\frac{\Delta m^{2}_{kj}L}{4E}}\Bigg]\,.
\end{align}
As mentioned, previous analyses considered the constant \ac{ULDM} limit, in which $\kappa_{ij}$ is strictly real, and given by Eq.~\eqref{eq:reKappaConstULDM}. In this case, one may interpret the first-order corrections as simple modifications to the PMNS matrix $U\rightarrow \tilde{U}$ as in Eq.~\eqref{eq:UtildeConst} and Eq.~\eqref{eq:TConst}. However, moving away from this limit, we may no longer use the constant $\tilde{U}$ description. This leads to the different energy dependence of the result associated with $\text{Im}\leri{\kappa}$ and with the non-zeroth order terms in $2 E m_\phi/\leri{\Delta m^2_{ij}}$ appearing in $\text{Re}\leri{\kappa}$\,.

\subsection{CPV effects}
\label{subsec:CPV}
\ac{CPV} is realized in vacuum in the three neutrino picture. The difference between the neutrino and antineutrino transition probability is given by
\begin{align}
\label{eq:3genCPVstandard}
\Delta P_{\alpha \beta}^{\text{standard}} & =4\sum_{i>j}\text{Im}\left(U_{\alpha i}^{*}U_{\beta i}U_{\alpha j}U_{\beta j}^{*}\right)\sin{\frac{\Delta m_{ij}^{2}L}{2E}}
=16 J \sin{\frac{\Delta m_{31}^{2}L}{4E}}\sin{\frac{\Delta m_{21}^{2}L}{4E}}\sin{\frac{\Delta m_{32}^{2}L}{4E}}\sum_\gamma \varepsilon_{\alpha\beta\gamma}\,,
\end{align}
where the $\varepsilon_{\alpha\beta\gamma}$ is the three-dimensional Levi-Civita tensor and $J$ is the Jarlskog invariant:
\begin{equation} 
J=\cos\theta_{13}\sin2\theta_{12}\sin2\theta_{13}\sin2\theta_{23}\sin\delta_{\text{CP}}.
\end{equation}
Notice that \ac{CPV} vanishes if we set any $\Delta m^2_{ij}$ to be zero, which will result in an effective two-neutrino picture. We therefore name this source of \ac{CPV} the ``$3-\nu$ \ac{CPV}". 

Let us now discuss \ac{CPV} in the presence of \ac{ULDM} interactions with neutrinos. From Eq.~\eqref{eq:P1alphabetaGeneric}, we find the \ac{CPV} part of neutrino probabilities, i.e. the difference between neutrinos and anti-neutrinos probabilities $\Delta P^{(1)}_{\alpha\beta}$ to be 
\begin{align}
\label{eq:DeltaP1}
\Delta P^{(1)}_{\alpha\beta}\equiv P^{(1)}_{\alpha\beta}-P^{(1)}_{\bar{\alpha}\bar{\beta}}&=-\frac{4 \phi_0}{ E}\sum_{i,j,k} \text{Im}\leri{U^*_{\beta k}U_{\alpha k}U_{\beta i}U^*_{\alpha j}{\tilde{m}_{ij}}}\text{Re}\leri{e^{i\frac{\leri{\Delta m^{2}_{jk}+\Delta m^{2}_{ik}}L}{4E}} {\kappa_{ij}}^*}\,.
\end{align}
Previous works have assumed the neutrino Hamiltonian to be constant throughout propagation, and thus the \ac{ULDM} effects were thought of as slow temporal modulations of the standard neutrino oscillation parameters. In this limit, as mentioned, $\kappa_{ij}$ is strictly real, and \ac{CPV} can be interpreted as a slow variation of the standard \ac{CPV} result~\cite{Losada:2021bxx}. Explicitly, in three generations, the contribution of the real part of $\kappa$ for $i\neq j\neq k$ and $\alpha\neq \beta$, for a specific $\hat{y}_{ij}=\hat{y}_{ji}\neq0$ is
\begin{align}
\label{eq:CPVRek}
\Delta P^{(1)}_{\alpha\beta}&=-8\frac{\phi_0}{ E} \text{Re}\leri{\kappa_{ij}}\sin{\frac{\Delta m^{2}_{jk}L}{4E}}\sin{\frac{\Delta m^{2}_{ik}L}{4E}}\text{Im}\leri{U^*_{\beta k}U_{\alpha k}\leri{U_{\beta i}U^*_{\alpha j}\tilde{m}_{ij}+U_{\beta j}U^*_{\alpha i}\tilde{m}^*_{ij}}}\,,
\end{align}
which yields in the $m_\phi L\rightarrow 0, 2Em_\phi/ \Delta m_{ij}^2\rightarrow 0 \,,  V\leri{t}\propto \sin{m_\phi t}$ limit
\begin{align}
\label{eq:3genCPVULDM}
\Delta P^{(1)}_{\alpha\beta}
\approx&-16\frac{\phi_0}{\Delta m^{2}_{ij}}\sin\leri{m_\phi \leri{t-\frac{L}{2}-t_0}}\cos{\frac{m_\phi L}{2}} \times\nonumber\\
&\times \sin{\frac{\Delta m^{2}_{ij}L}{4 E}}\sin{\frac{\Delta m^{2}_{jk}L}{4E}}\sin{\frac{\Delta m^{2}_{ik}L}{4E}}\text{Im}\leri{U^*_{\beta k}U_{\alpha k}\leri{U_{\beta i}U^*_{\alpha j}\tilde{m}_{ij}+U_{\beta j}U^*_{\alpha i}\tilde{m}^*_{ij}}}
\,.
\end{align}
Since we keep only first-order terms in the \ac{ULDM} couplings, which here we assume to be off-diagonal in the \ac{ULDM}-independent mass basis, one may interpret this result as the effect of the first-order correction to the standard Jarlskog invariant.

For diagonal couplings $\hat{y}_{ii}$, both $\tilde{m}_{ii}$ and $\kappa_{ii}$ are real, and thus
\begin{align}
\Delta P^{(1)}_{\alpha\beta}\leri{t}
&=-\frac{4\tilde{m}_{ii} \phi_0}{ E m_\phi}\sin\leri{m_\phi \leri{t-\frac{L}{2}-t_0}}\sin\leri{\frac{m_\phi L}{2}}\times\nonumber\\
&\times \text{Im}\leri{U_{\beta j}U^*_{\alpha j}U^*_{\beta i}U_{\alpha i}}\sin{\leri{\frac{\leri{\Delta m^{2}_{ik}+\Delta m^{2}_{ij}}L}{4E}}}\sin{\leri{\frac{\Delta m^{2}_{kj}L}{4E}}}\,.
\end{align}
This \ac{CPV} term is of course proportional to $J$, as only the real part of diagonal couplings affects the neutrino Hamiltonian. This term is simply a correction to the mass differences of the neutrinos coming from the \ac{ULDM} coupling, as shown in Eq.~\eqref{eq:diagonalDeltam}.

For off-diagonal couplings, stepping away from the $m_\phi \rightarrow 0$ limit, we should also consider the contribution of the imaginary part of $\kappa$
\begin{align}
\label{eq:CPVImk}
\Delta P^{(1)}_{\alpha\beta}=& 4\text{Im}\leri{\kappa_{ij}}\sum_{i,j,k} \text{Im}\leri{U^*_{\beta k}U_{\alpha k}U_{\beta i}U^*_{\alpha j}{\tilde{m}_{ij}}}\sin\leri{\frac{\leri{\Delta m^{2}_{jk}+\Delta m^{2}_{ik}}L}{4E}}\,.
\end{align}
For $i\neq j\neq k$ and $\alpha\neq \beta$, for a specific $\hat{y}_{ij}=\hat{y}_{ji}\neq0$, it becomes
\begin{align}
\Delta P^{(1)}_{\alpha\beta}=8\text{Im}\leri{\kappa_{ij}}\Big[&\text{Im}\leri{U^*_{\beta i}U_{\alpha i}\leri{U_{\beta j}U^*_{\alpha i}{\tilde{m}^*_{ij}}-U_{\beta i}U^*_{\alpha j}{\tilde{m}_{ij}}}}\sin{\frac{\Delta m^{2}_{ik}L}{4E}}\cos{\frac{\Delta m^{2}_{jk}L}{4E}}\nonumber\\
+&\text{Im}\leri{U^*_{\beta j}U_{\alpha j}\leri{U_{\beta j}U^*_{\alpha i}{\tilde{m}^*_{ij}}-U_{\beta i}U^*_{\alpha j}{\tilde{m}_{ij}}}}\cos{\frac{\Delta m^{2}_{ik}L}{4E}}\sin{\frac{\Delta m^{2}_{jk}L}{4E}}\Big]\,.\label{eq:CPVImkTransition}
\end{align}
This expression is fundamentally different from the ``3-$\nu$ \ac{CPV}" effect, as it ``contains" the 2-$\nu$ \ac{CPV} effect we discussed in Section~\ref{sec:2_nu_CPV} (see Eq.~\eqref{eq:2nuCPV}). This can be seen in three distinct ways, all pointing to cases in which there would be no \ac{CPV} in the constant \ac{ULDM} potential limit. 

First, note that Eq.~\eqref{eq:3genCPVULDM} vanishes for $U^*_{\beta k}U_{\alpha k}=0$\,, namely, when the projection of the mass eigenstate not coupled to the \ac{ULDM} on either the outgoing or incoming interaction eigenstates vanishes. This case corresponds to an effective $J=0$. However, taking this limit for Eq.~\eqref{eq:CPVImkTransition}\,, we obtain
\begin{align}
\Delta P^{(1)}_{\alpha\beta} =&8\text{Im}\leri{\kappa_{ij}}\text{Im}\leri{U^*_{\beta i}U_{\alpha i}\leri{U_{\beta j}U^*_{\alpha i}{\tilde{m}^*_{ij}}-U_{\beta i}U^*_{\alpha j}{\tilde{m}_{ij}}}}\sin{\frac{\Delta m^{2}_{ij}L}{4E}}\,.
\end{align}
This expression does not vanish generically. Specifically, if say only $U_{\alpha k}=0$\,, then unitarity implies $|U_{\alpha i}|^2+|U_{\alpha j}|^2=1$\,, and we may denote $|U_{\alpha i}|^2\equiv \sin^2{\theta_\alpha}$ and obtain
\begin{align}
\Delta P^{(1)}_{\alpha\beta} =&16\cos{2\theta_\alpha}\text{Im}\leri{\kappa_{ij}}\text{Im}\leri{U_{\beta i}U^*_{\beta j}{\tilde{m}_{ij}}}\sin{\frac{\Delta m^{2}_{ij}L}{4E}}\,.
\end{align}
This effect is indeed the result of the two-generations-like effect, and thus does not vanish. 

Second, as explained above, we expect the standard \ac{CPV} effect in three generations to vanish when any of the mass differences $\Delta m^2_{ij}$ vanishes. Indeed, the standard \ac{CPV} in Eq.~\eqref{eq:3genCPVstandard} vanishes at this limit, as well as the constant-\ac{ULDM} \ac{CPV} in Eq.~\eqref{eq:3genCPVULDM} when considering a specific off-diagonal $\hat{y}_{ij}$\,, and setting $\Delta m^2_{jk}=0$, for $k\neq i\neq j$. However, at this limit, Eq.~\eqref{eq:CPVImkTransition} yields a non-zero result, and reads
\begin{align}
\Delta P^{(1)}_{\alpha\beta}=&8\text{Im}\leri{\kappa_{ij}}\text{Im}\leri{U^*_{\beta i}U_{\alpha i}\leri{U_{\beta j}U^*_{\alpha i}{\tilde{m}^*_{ij}}-U_{\beta i}U^*_{\alpha j}{\tilde{m}_{ij}}}}\sin{\frac{\Delta m^{2}_{ij}L}{4E}}\,,\label{eq:CPVImkTransitionDeltam}
\end{align}
exactly as in the $U^*_{\beta k}U_{\alpha k}=0$ case (however here we do not assume a specific structure for $U$). Given the known hierarchy in the neutrino mass differences, $\Delta m^2_{12}\ll\Delta m^2_{31},\Delta m^2_{32}$, we then expect that the new 2-$\nu$ \ac{CPV} effect resulting from $\hat{y}_{31}$ or $\hat{y}_{32}$ would dominate over the constant \ac{ULDM} \ac{CPV} roughly for $m_\phi\gtrsim \Delta m^2_{12}/\leri{2E}$. Note that if the neutrinos coupled to the \ac{ULDM} are degenerate in the unperturbed system, i.e. $\hat{y}_{ij}\neq0$ for $\Delta m^2_{ij}L/\leri{4E}\rightarrow 0$, the constant \ac{ULDM} limit $m_\phi L\rightarrow 0$ yields a non-zero \ac{CPV}, as $\text{Re}\leri{\kappa_{ij}}\rightarrow \sin\leri{m_\phi\leri{t-t_0}} L/2$. This corresponds to a leading order effective mass difference induced by the \ac{ULDM} potential. In this limit, we obtain $\text{Im}\leri{\kappa_{ij}}\propto L^3 m_\phi \Delta m^2_{ij}/\leri{4E}$\,, and thus the 2-$\nu$ \ac{CPV} effect is expected to be sub-dominant to the constant \ac{ULDM} effect. 

Finally, note that standard \ac{CPV} does not affect survival probabilities. Accordingly, the constant-\ac{ULDM} \ac{CPV} in Eq.~\eqref{eq:3genCPVULDM} is also zero for $\alpha=\beta$. The new \ac{CPV} effect we found in Eq.~\eqref{eq:CPVImk} for $\alpha=\beta$ does not vanish generically, and is given for a specific $\hat{y}_{ij}=\hat{y}_{ji}\neq0$ with $i\neq j\neq k$ by 
\begin{align}
\label{eq:CPVImksurvival}
\Delta P^{(1)}_{\alpha\alpha}
=&-8\text{Im}\leri{\kappa_{ij}}\text{Im}\leri{\tilde{m}_{ij}U_{\alpha i}U^*_{\alpha j}}\times\nonumber\\
\times&\leri{\leri{2|U_{\alpha j}|^2+1}\sin\leri{\frac{\Delta m^{2}_{jk}L}{4E}}\cos\leri{\frac{\Delta m^{2}_{ik}L}{4E}}+\leri{2|U_{\alpha i}|^2+1}\cos\leri{\frac{\Delta m^{2}_{jk}L}{4E}}\sin\leri{\frac{\Delta m^{2}_{ik}L}{4E}}}\,.
\end{align}
Therefore, even experiments only measuring survival probabilities would be sensitive to \ac{CPV} \ac{ULDM}-neutrino couplings. This is a completely new prediction resulting from our analysis, considering the time variations of the Hamiltonian within the neutrino propagation time.

\section{Experimental implications - Rayleigh Periodogram} \label{sec: Rayleigh}
In Section~\ref{sec:analytic_expressions}, we introduced analytical expressions for the first-order modifications to neutrino probabilities resulting from \ac{ULDM}-neutrino interactions. We are interested in devising a method for detecting these modifications in neutrino oscillations experiments. As the effects we found are time dependent, characterized by simple oscillatory functions with periods $\tau_\phi=2\pi/m_\phi$, we are essentially interested in measuring the spectral power of the neutrino probabilities at an angular frequency corresponding to $m_\phi$. 

Consider the following time scales relevant for neutrino oscillation experiments:
\begin{enumerate}
	\item $\tau_{e}$: The running time of the experiment.
	\item $\tau_{s}=\frac{\tau_{e}}{N_{\nu}}$: The average spacing between
	events, where $N_{\nu}$ is the total number of measured events.
	\item $\tau_{r}$: The resolution of the event timing. While in practice the effective resolution could be set by different sources of temporal uncertainty (e.g. beam spread), we will refer to it as the ``clock" resolution. 
\end{enumerate}
We notice the following hierarchy between the time scales:
\begin{equation}
\tau_{r}\ll\tau_{s}\ll\tau_{e}.
\end{equation}
Since the frequencies we are interested in are determined by $\tau_\phi$, one would naively assume that the experimental sensitivity to \ac{ULDM} masses $m_\phi\gtrsim 2/\tau_s$ would be suppressed, as the statistical uncertainty on the neutrino probability binned over the corresponding $\tau_\phi$ would be quite large, due to a small number of events expected to occur within that period. However, as we will now show, the true limiting factor, in terms of the experimental sensitivity, is rather the effective clock resolution $\tau_r$, which is much smaller than $\tau_s$. 

To understand this, first recall that the events are inherently binned over times $\tau_r$, while $\tau_s$ is not associated with some instrumental resolution, and binning over it is completely artificial, and is done post-measurement. Let us then use the full data, given as the number of neutrinos of a certain flavor detected at a sequence of $N_t=\tau_e/\tau_r$ clock ticks, each with short duration $\tau_r$. Since $\tau_r\ll \tau_s$, we ignore the possibility that two neutrinos might be detected during $\tau_r$, and define the following power spectrum:
\begin{align}
z\left(f\right)=\frac{2}{N_\nu}\left(\left[\sum_{n=1}^{N_{t}}h\left(t_{n}\right)\cos\left(2\pi f t_{n}\right)\right]^{2}+\left[\sum_{n=1}^{N_{t}}h\left(t_{n}\right)\sin\left(2\pi f t_{n}\right)\right]^{2}\right)\,,
\label{eq:almostRayleigh}
\end{align}
with
\begin{equation}
h\left(t_{i}\right)=\left\{ \begin{matrix}1 & \text{event detected,}\\
0 & \text{no event detected.}
\end{matrix}\right.\label{eq:h(ti)}
\end{equation}
This definition gives us the well known Rayleigh power spectrum
\begin{equation}
z\left(f\right)=\frac{2}{N_\nu}\left(\left[\sum_{n=1}^{N_{\nu}}\cos\left(2\pi f t_{n}\right)\right]^{2}+\left[\sum_{n=1}^{N_{\nu}}\sin\left(2\pi f t_{n}\right)\right]^{2}\right).
\label{eq:Rayleigh}
\end{equation}
Notice that the summation now is over the neutrino events instead of over time bins. For a list of events detected in a time series $\{t_{n}\}$, we calculate the power spectrum $z$, which
is an analytic function of the probed frequencies $f$. For $\{t_{n}\}$
uniformly random distributed between $0<t_{n}<\tau_{e}$, the sum of sines and sum of cosines in~Eq.~\eqref{eq:Rayleigh} are each normally distributed with mean value of zero and standard deviation of $\sqrt{N_\nu/2}$. Thus $z(f)$, being their appropriately normalized sum of squares, is Chi-squared distributed with two degrees of freedom, which is simply an exponential distribution of $z$. This means, under the background only hypothesis, that if we consider a specific
frequency $f'$ and calculate its power $z\left(f'\right)$, its probability
to be higher than some value $Z$ is independent of the frequency $f'$ (assuming $f'\gtrsim 1/\tau_e$ and no other spectral noise sources), and given by 
\begin{equation}
p\left(z>Z\right)=\int\limits_{Z}^{\infty} \frac12 e^{-z/2} dz=e^{-Z/2}.
\end{equation}

Let us estimate the signal in case of a temporal modulation of the neutrino probability. The probability of detecting a neutrino of some flavor at time $t_i$ in an energy bin $[E_{j,\min},E_{j,\max}]$ is given by 
\begin{equation}
P\left(h\left(t_{i},E_{j}\right)=1\right)=\int\limits _{E_{j,\min}}^{E_{j,\max}}\int\limits _{t_{i}-\tau_{r}/2}^{t_{i}+\tau_{r}/2}dtdEF\left(E\right)P\left(E,t\right),\label{eq:P(h)OneBin}
\end{equation}
where $F$ is the unoscillated neutrino flux, i.e. the flux that would have reached the detector had all the produced neutrinos oscillated into the flavor in question
\begin{align}
F\left(E\right) & =\frac{dN_{\text{unoscillated}}}{dEdt},
\end{align}
and $P\left(E,t\right)$ is the oscillation probability given by 
\begin{align}
P\left(E,t\right) & =P^{\left(0\right)}\left(E\right)+P^{\left(1\right)}\left(E,t\right)\,,
\end{align}
with $P^{(0)}$, $P^{(1)}$ given in Eqs.~(\ref{eq:P0}) and (\ref{eq:P1}).
The total number of events collected from different energy bins over the entire duration of the experiment $\tau_e$, assuming the \ac{ULDM} modification to the probability is very small, is then 
\begin{align}
N_{\nu} &
\approx \int\limits _{E_{\min}}^{E_{\max}}\int\limits _{0}^{\tau_{e}}dtdEF\left(E\right)P^{\left(0\right)}\left(E\right)
=\tau_{e}\int\limits _{E_{\min}}^{E_{\max}}dEF\left(E\right)P^{\left(0\right)}\left(E\right)\,.
\end{align}
Similarly, one may define the corresponding number of events associated with the small \ac{ULDM} probability
\begin{align}
N_{\nu,s}^{\left(1\right)} & =\tau_{e}\int\limits _{E_{\min}}^{E_{\max}}dEF\left(E\right)P_{s}^{\left(1\right)}\left(E\right)\,,\\
N_{\nu,c}^{\left(1\right)} & =\tau_{e}\int\limits _{E_{\min}}^{E_{\max}}dEF\left(E\right)P_{c}^{\left(1\right)}\left(E\right)\,,
\end{align}
where $P_s$ and $P_c$ can be read off from
\begin{equation}
P^{\left(1\right)}\left(E,t\right)=P_{s}^{\left(1\right)}(E)\sin\left(m_{\phi}t\right)+P_{c}^{\left(1\right)}(E)\cos\left(m_{\phi}t\right)\,.
\end{equation}
To identify $P_{s}^{\left(1\right)}(E)$ and $P_{c}^{\left(1\right)}(E)$, note that the imaginary and real components of $\kappa_{ij}$ determine the time dependence of $P^{(1)}$, as can be seen in Eqs. (\ref{eq:Rek}) and (\ref{eq:Imk}). Since $\text{Re}(\kappa)$ and $\text{Im}(\kappa)$ are orthogonal in time, but have the same frequency, $P^{(1)}$ can be written as
\begin{align}
P^{(1)} = a\sin(m_\phi(t-L/2-t_0))+b\cos(m_\phi(t-L/2-t_0))\,,
\end{align}
and thus 
\begin{align}
P_{s}^{\left(1\right)} &= a\cos\leri{m_\phi\leri{L/2+t_0}}+b\sin\leri{m_\phi\leri{L/2+t_0}}\,,\\
 P_{c}^{\left(1\right)} &=-a\sin\leri{m_\phi\leri{L/2+t_0}}+b\cos\leri{m_\phi\leri{L/2+t_0}}\,.
\end{align}
We may now calculate the expected value of $z(f)$ at $f=m_\phi/\leri{2\pi}$ in the presence of a \ac{ULDM}, with mass $m_\phi$, interacting with neutrinos\footnote{ We do not specify the range and resolution of frequencies at which a peak in the Rayleigh spectrum is searched for, as there is no clear recipe for determining them. One can scan frequencies as small as $\sim\tau_e^{-1}$, and as large as $\tau_r^{-1}$, since the effect is suppressed for frequencies higher than that. Typically, the resolution of the scan would be approximately $\sim\tau_e^{-1}$, as this is the width of the features in $z(f)$. However, the authors of \cite{SNO:2009ktr}, for example, chose to oversample with a higher resolution. This comes at the cost of a more significant look-elsewhere effect \cite{Gross:2010qma}, which reduces the confidence level of the signal. Since this effect is mainly determined by the number of frequencies one scans, a number which depends neither on the experiment nor on the \ac{ULDM} parameters, we did not include it in the analyses in this paper.}. One can show, see Appendix \ref{appendix: Rayleigh}, that for $m_\phi \lesssim 2/\tau_r$ 
\begin{align}
\left\langle \left[\sum_{i=1}^{N_{\nu}}\sin\left(2\pi t_{i}f\right)\right]^{2}\right\rangle  & =\frac{N_{\nu}}{2}+\frac{\left(N_{s}^{\left(1\right)}\right)^{2}}{4}\,,\label{eq:sines}\\
\left\langle \left[\sum_{i=1}^{N_{\nu}}\cos\left(2\pi t_{i}f\right)\right]^{2}\right\rangle  & =\frac{N_{\nu}}{2}+\frac{\left(N_{c}^{\left(1\right)}\right)^{2}}{4}\,,\label{eq:cosines}
\end{align}
and thus 
\begin{align}
\label{eq:z_Mean}
\left\langle z\left(f\right)\right\rangle  & =\frac{2}{N_{\nu}}\left[\left\langle \left[\sum_{i=1}^{N_{\nu}}\sin\left(2\pi t_{i}f\right)\right]^{2}\right\rangle +\left\langle \left[\sum_{i=1}^{N_{\nu}}\cos\left(2\pi t_{i}f\right)\right]^{2}\right\rangle \right] =2+\frac{\left(N_{s}^{\left(1\right)}\right)^{2}+\left(N_{c}^{\left(1\right)}\right)^{2}}{2N_\nu}\,.
\end{align} 

Note a few key points regarding the above result: 
\begin{itemize}
	\item The expected sensitivity to a \ac{ULDM} coupling $\hat{y}$ improves with the square-root of the (unoscillated) neutrino flux of the experiment $\sqrt{F}$, and decreases as the square-root of the expected vacuum oscillation probability $\sqrt{P^{(0)}}$. 
	\item Assuming the \ac{ULDM} perturbation is coherent throughout the experimental time\,, the sensitivity to $\hat{y}$ improves as $\sqrt{\tau_e}$. For a coherence time $\tau_\text{c}$ satisfying $L,\tau_r<\tau_\text{c}< \tau_e$, the sensitivity to $\hat{y}$ would follow $ \sqrt{\tau_e}\leri{\frac{\tau_c}{\tau_e}}^\frac{1}{4}$. While our results were derived assuming $\tau_c\gtrsim \tau_r,L$, it is possible to derive the appropriate expressions for $\tau_c<L,\tau_r$ by statistically averaging over the \ac{ULDM} phase (either in determining $P^{(1)}$ from the Dyson integral over $L$ in Eq.~\eqref{eq:U1Dyson}, or in the integral over $\tau_r$ in Eq.~\eqref{eq:P(h)OneBin}), which might significantly suppress the linear contributions to the neutrino probabilities.
	\item As stated, since the distribution of $z\leri{f}$ under the null hypothesis is frequency independent, the expected sensitivity of the experiment for different \ac{ULDM} masses is completely determined by the mass dependence of $P^{(1)}$, assuming $m_\phi\lesssim 2/\tau_r$ and $\tau_c\gtrsim \tau_e$ (or mass-independent coherence times). If the \ac{ULDM} coherence time is determined by $\tau_c = 2\pi/\leri{m_\phi \beta^2}$ with the virial velocity $\beta\approx 10^{-3}$\,, then for $m_\phi > 2\pi/\leri{\tau_e \beta^2}$ one can think of $P^{(1)}$ as being effectively suppressed by an additional factor of $\leri{\frac{2\pi}{\tau_e m_\phi \beta^2}}^\frac{1}{4}$.
	For $m_\phi> 2/\tau_r$, the integration over time $\tau_r$ in Eq.~\eqref{eq:P(h)OneBin} would yield an effective additional $2/\leri{m_\phi \tau_r}$ suppression to $P^{(1)}$. Also note that our result coincides with the result obtained by previous analyses assuming $\tau_s<\tau_\phi$ if one sets $\tau_r=\tau_s$. Therefore, compared to previous analyses, the Rayleigh method would yield bounds on the couplings of \ac{ULDM} of masses $2/\tau_r>m_\phi> 2/\tau_s$ that are stronger by $m_\phi \tau_s$, and bounds that are stronger by $\tau_s/\tau_r$ for $m_\phi>2/\tau_r$. 
	\item While the \ac{ULDM} phase (denoted by $m_\phi t_0$) affects the sum of cosines and the sum of sines in Eq~\eqref{eq:cosines} and Eq.~\eqref{eq:sines}, respectively, it does not affect $z(f)$, as they are added in quadrature. Namely, the calculation of $z(f)$ does not require a knowledge of the \ac{ULDM} phase, and the phase of the cosines and sines used for it may be chosen arbitrarily. Consequently, in order to calculate the expected bounds, one could treat the time-independent factors associated with $\text{Re}\leri{\kappa_{ij}}$ and $\text{Im}\leri{\kappa_{ij}}$ separately, as they will be added in quadrature. Of course, scanning over the phase and examining the sum of cosines and sum of sines independently would allow recovering the \ac{ULDM} phase information.
	
	\item While in our calculation we used the simple modulation function $h(t_i)$ in Eq.~\eqref{eq:h(ti)}, one may use other choices for $h$ in order to isolate or amplify specific properties of the signal. Specifically, assigning $h\leri{t_i}=h\leri{E}$ would effectively set a different, energy-dependent, weight for the neutrino events, which could be helpful, for example, in recovering energy-dependent highly oscillatory components of the signal. Another useful choice could be made to isolate either the \ac{CPC} or the \ac{CPV} probabilities in an experiment that allows for detecting both neutrinos and anti-neutrinos, as
	\begin{equation}
	h_{\pm}\left(t_{i}\right)=\left\{ \begin{matrix}1/n_{\nu} & \nu\text{ event detected,}\\
	\pm 1/n_{\bar{\nu}} & \overline{\nu}\text{ event detected,}\\
	0 & \text{no event detected\,,}
	\end{matrix}\right.
	\end{equation}
	where $n_{\nu}$ is the total number of neutrinos detected and $n_{\bar{\nu}}$ is the total number of anti-neutrinos detected. Generally speaking, for any choice of a modulation function $h(\phi)$ which is time-independent (let it be a function of the neutrino energy, flavor, lepton number etc.), we may define
	\begin{align}
	\tilde{r}\leri{f}=\leri{\left[\sum_{i=1}^{N_{\nu}}h\leri{{\phi}_i}\sin\left(2\pi t_{i}f\right)\right]^{2}+\left[\sum_{i=1}^{N_{\nu}}h\leri{\phi_i}\cos\left(2\pi t_{i}f\right)\right]^{2}}\,.
	\end{align}
	Then, the modified Rayleigh score 
	\begin{align}
	\tilde{z}\leri{f}=\frac{2}{\tilde{N}^0_\nu}\tilde{r}\leri{f}\,,
	\end{align}
	is $\chi_2^2$ distributed, and in the presence of a signal its expectation value is
	\begin{align}
	\langle \tilde{z}\leri{f}\rangle =2+\frac{\left(\tilde{N}_{s}^{\left(1\right)}\right)^{2}+\left(\tilde{N}_{c}^{\left(1\right)}\right)^{2}}{2\tilde{N}^0_\nu}\,,
	\end{align}
	with
	\begin{align}
	\tilde{N}^0_\nu &=\left \langle \sum_{\rm bkg} h^2 \right\rangle  =\tau_{e}\int\limits _{E_{\min}}^{E_{\max}}\int_\phi\frac{dN}{dE dt d\phi} P^{\left(0\right)} h^2 d\phi dE\,,\\
	\tilde{N}_{\nu,s}^{\left(1\right)} &= \left \langle \sum_{\rm s} h \right \rangle  =\tau_{e}\int\limits _{E_{\min}}^{E_{\max}}\int_\phi\frac{dN}{dE dt d\phi} P_{s}^{\left(1\right)} h d\phi dE\,,\\
	\tilde{N}_{\nu,c}^{\left(1\right)} & =\left \langle \sum_{\rm c} h \right \rangle =\tau_{e}\int\limits _{E_{\min}}^{E_{\max}}\int_\phi\frac{dN}{dE dt d\phi} P_{c}^{\left(1\right)}h d\phi dE\,.
	\end{align}
	Then, concretely, for $h_{\pm}$ one would obtain
	\begin{align}
	\langle \tilde{z}\leri{f}\rangle_{\pm} \approx 2+\frac{\left(\epsilon_{s}^{\pm}\right)^{2}+\left(\epsilon_{c}^{\pm}\right)^{2}}{2\leri{\frac{1}{n_\nu}+\frac{1}{n_{\bar{\nu}}}}}\,,
	\end{align}
	with
	\begin{align}
	\epsilon_{s,c}^{+}&= \frac{\int\limits _{E_{\min}}^{E_{\max}}dEF\left(E\right)2\Sigma P_{s,c}^{\left(1\right)}\left(E\right)}{\int\limits _{E_{\min}}^{E_{\max}}dEF\left(E\right) P^{\left(0\right)}\left(E\right)}\,,\\
	\epsilon_{s,c}^{-}&= \frac{\int\limits _{E_{\min}}^{E_{\max}}dEF\left(E\right)\Delta P_{s,c}^{\left(1\right)}\left(E\right)}{\int\limits _{E_{\min}}^{E_{\max}}dEF\left(E\right) P^{\left(0\right)}\left(E\right)}\,,
	\end{align}	
	with $\Sigma P^{\left(1\right)}$ and $\Delta P^{\left(1\right)}$ defined in Eq.~\eqref{eq:SigmaP1} and Eq.~\eqref{eq:DeltaP1}, respectively.
\end{itemize}

\section{Current and future-projected bounds}
\label{sec:bounds}
In this section we study the sensitivity of various neutrino oscillation experiments to the \ac{ULDM} modulation amplitude using the Rayleigh periodogram. We derive bounds on the neutrino-\ac{ULDM} couplings from existing experiments which have already looked for time modulations in neutrino survival and transition probabilities, and projected sensitivities for future experiments. We do not use Monte-Carlo simulations, but use the scaling that we derived for the expectation value of the magnitude of the Rayleigh spectrum peak at the modulation frequency in Eq.~\eqref{eq:z_Mean}, with the parameters of the experiment and of the model. 
When calculating the height of the peak, we assumed that the \ac{ULDM} modulation amplitude is predominantly affected by one entry of the $\hat{y}$ matrix at a time. We also provide a separate analysis for the real and imaginary components of the off-diagonal entries of $\hat{y}$, assuming $\delta_{CP}=0$. Note that in our calculations we assumed a normal hierarchy, and that $m_1=0$. The latter assumption can be relaxed by re-identifying $\text{Im}\leri{y_{1i}}\rightarrow \leri{m_i-m_1}\text{Im}\leri{y_{1i}}/m_i$\,, $\text{Re}\leri{y_{1i}}\rightarrow \leri{m_i+m_1}\text{Re}\leri{y_{1i}}/m_i$ for the off-diagonal couplings, and $y_{22}\rightarrow y_{22}-m_1y_{11}/m_2\,, y_{33}\rightarrow y_{33}-m_1y_{11}/m_3$ for the diagonal ones. For the projected sensitivities, we find the curve in parameter space $\hat{y}\leri{m_\phi}$ for which the peak is expected to match the 95\% percentile of the $\chi^2_2$ distribution, characterizing the background (corresponding to a CL=0.95). The corresponding bounds in terms of the \ac{ULDM} parameters are shown in Fig.~\ref{fig: Experiments}.

\subsection{Current bounds}
\subsubsection{Daya-Bay}
The Daya Bay experiment collects $\bar{\nu}_e$ events from nearby reactors which are approximately $800$ meters away from the detector. The collaboration has searched for time-dependent modulations of the survival probability $P_{\bar{e}\bar{e}}$, for frequencies between $5.9\times 10^{-5}$/sidereal hour to 0.5/sidereal hour, and found no significant evidence \cite{DayaBay:2018fsh}. We plot in Fig.~\ref{fig: Experiments} the \ac{ULDM} couplings corresponding to an expected sensitivity at CL=0.95, assuming 621 days of data taking, with a total of 800 unoscillated events per day \cite{DayaBay:2016ggj}.

\subsubsection{Solar experiments - SNO}
The Super-K and SNO experiments collect solar $\nu_e$ with event rates of $15$/day \cite{Super-Kamiokande:2003snd} and $10$/day \cite{Tolich:2011zza}, respectively. Due to the MSW effect, neutrinos with energy larger than $\sim 1$ MeV leave the sun as the $\nu_2$ mass states. Therefore, their survival probability does not oscillate with propagation, and is given by $P^0_{2e}=|U_{e2}|^2\approx\sin^2{\theta_{12}}$. This means that up to day/night matter effects, the solar neutrino flux should be constant along the year. 

To calculate the first-order correction coming from the \ac{ULDM}, we assume that $m_\phi \tau_{\odot} \ll1$, where $\tau_\odot$ is the longest time-scale for neutrino effects in the sun, such that the neutrino leaves the sun as an eigenstate of the instantaneous Hamiltonian, $\mathcal{H}(t_i)\approx \mathcal{H}_0+\tilde{m}\sin\leri{m_\phi \leri{t_i-t_0}}$, denoted by $\nu_m\leri{t_i}$. We then turn to our previously obtained expressions for $P_{\alpha \beta}$, noting that $\nu_\alpha=\nu_m\leri{t_i}$ with $m=2$. Namely, the PMNS-like matrix $U_{\alpha i}$ should be replaced by the projection of the instantaneous mass eigenstate at time $t_i$ onto the \ac{ULDM}-independent mass eigenstates - $U_{\alpha i}\rightarrow T_{2 i}\leri{t_i}$\, where $T(t_i)$ is the diagonalizing matrix of $\mathcal{H}(t_i)$ in the \ac{ULDM}-independent mass basis. The deviation of $T(t_i)$ from the identity yields a first order correction in the \ac{ULDM} coupling to $\mathcal{U}^{(0)}$. When calculating $\mathcal{U}^{(1)}$, one may set $U_{\alpha i}=\delta_{2i}$, since it is already linear in the \ac{ULDM} potential. For off-diagonal \ac{ULDM} couplings $\hat{y}_{i2}$ with $i\neq 2$, the leading \ac{ULDM} correction in our formalism is given by
\begin{align}
P^{(1)}_{2e}& =2\frac{\phi_0}{E}\text{Re}\leri{U^*_{e 2}U_{e i}\tilde{m}_{i2}}\leri{-\frac{E\sin\leri{m_\phi \leri{t-L-t_0}}}{\Delta m^2_{i2}}\cos{\frac{\Delta m^{2}_{i2}L}{2E}}-\cos{\frac{\Delta m^{2}_{i2}L}{4E}}\text{Im}\leri{\kappa_{i2}}+\sin{\frac{\Delta m^{2}_{i2}L}{4E}}\text{Re}\leri{\kappa_{i2}}}\nonumber\\
&-2\frac{\phi_0}{E}\text{Im}\leri{U^*_{e 2}U_{e i}\tilde{m}_{i2}}\leri{\frac{E\sin\leri{m_\phi \leri{t-L-t_0}}}{\Delta m^2_{i2}}\sin{\frac{\Delta m^{2}_{i2}L}{2E}}+\cos{\frac{\Delta m^{2}_{i2}L}{4E}}\text{Re}\leri{\kappa_{i2}}+\sin{\frac{\Delta m^{2}_{i2}L}{4E}}\text{Im}\leri{\kappa_{i2}}}\,,\label{eq:Psolar}
\end{align}
whereas for the diagonal coupling $\hat{y}_{22}$ and for all $\hat{y}_{ij}$ with $j\neq2$\,, there are no first-order contributions to the solar probability.

The SNO collaboration looked for a time variation in the solar neutrino flux with periods between $10$ minutes to $1$ day, and found a null result with confidence level of $90\%$~\cite{SNO:2009ktr}. The search was expected to be sensitive (have a 90\% probability of making a CL=0.99 discovery) to sinusoidal modulations at these frequencies with an amplitude of 0.12 or greater, compared to the time-averaged flux. We may then interpret this result as a bound on the \ac{ULDM} parameters, assuming $m_\phi$ is equal to the modulation frequency. In the $m_\phi\ll \frac{\Delta m^2_{i2} }{2 E}$ limit, which applies throughout the frequency range of the analysis of~\cite{SNO:2009ktr}, the correction to the solar probability in Eq.~\eqref{eq:Psolar} becomes 
\begin{align}
P^{(1)}_{2e}& \approx \frac{2\phi_0}{\Delta m^2_{i2}}\text{Re}\leri{U_{e 2}U^*_{e i}\tilde{m}_{i2}^*}\Big(-\sin{\leri{m_\phi \leri{t-t_0}}}+\frac{2E m_\phi}{\Delta m^2_{i2}}\sin{\frac{\Delta m^{2}_{i2}L}{2E}}\cos\leri{m_\phi \leri{t-L-t_0}}\Big)\nonumber\\
&+4\frac{ E m_\phi\phi_0}{{\Delta m^2_{i2}}^2}\text{Im}\leri{U_{e 2}U^*_{e i}\tilde{m}_{i2}^*}\Big(-\cos\leri{m_\phi \leri{t-t_0}}+\cos{\frac{\Delta m^{2}_{i2}L}{2E}}\cos\leri{m_\phi \leri{t-L-t_0}}\Big)\,.\label{eq:PsolarLowMasses}
\end{align}

\subsection{Projected sensitivities}
\subsubsection{Reactor experiments}
In addition to current bounds from the Daya-Bay experiment, we study the projected sensitivities of \ac{ULDM} parameters from other reactor neutrino experiments. We consider JUNO that measures the survival probability of $\bar{\nu}_e$ that travel $\sim53$ km. The energy resolution and baseline length of JUNO enable it to be sensitive to both neutrino mass squared differences $\Delta m^2_{21}$, $\Delta m^2_{31}$ and thus to determine the neutrino mass hierarchy. It has an expected flux of $83$ unoscillated events per day \cite{Giaz:2018gdd}, and its projected sensitivity to \ac{ULDM} parameters is presented in Fig.~\ref{fig: Experiments}.

We did not consider the KamLAND experiment in our analysis, as due to its long baseline of $\sim180$ km has a relatively low flux of $1-2$ events per day. As opposed to accelerator experiments which fire a narrow beam of neutrinos, reactor experiments fire neutrinos uniformly in every direction and therefore their flux decays with $L^{-2}$. It also collects events from different reactors with different distances, which considerably complicates the analysis.
\subsubsection{Accelerator experiments}
\begin{figure}[h]
	\begin{centering}
		\includegraphics[scale=0.45]{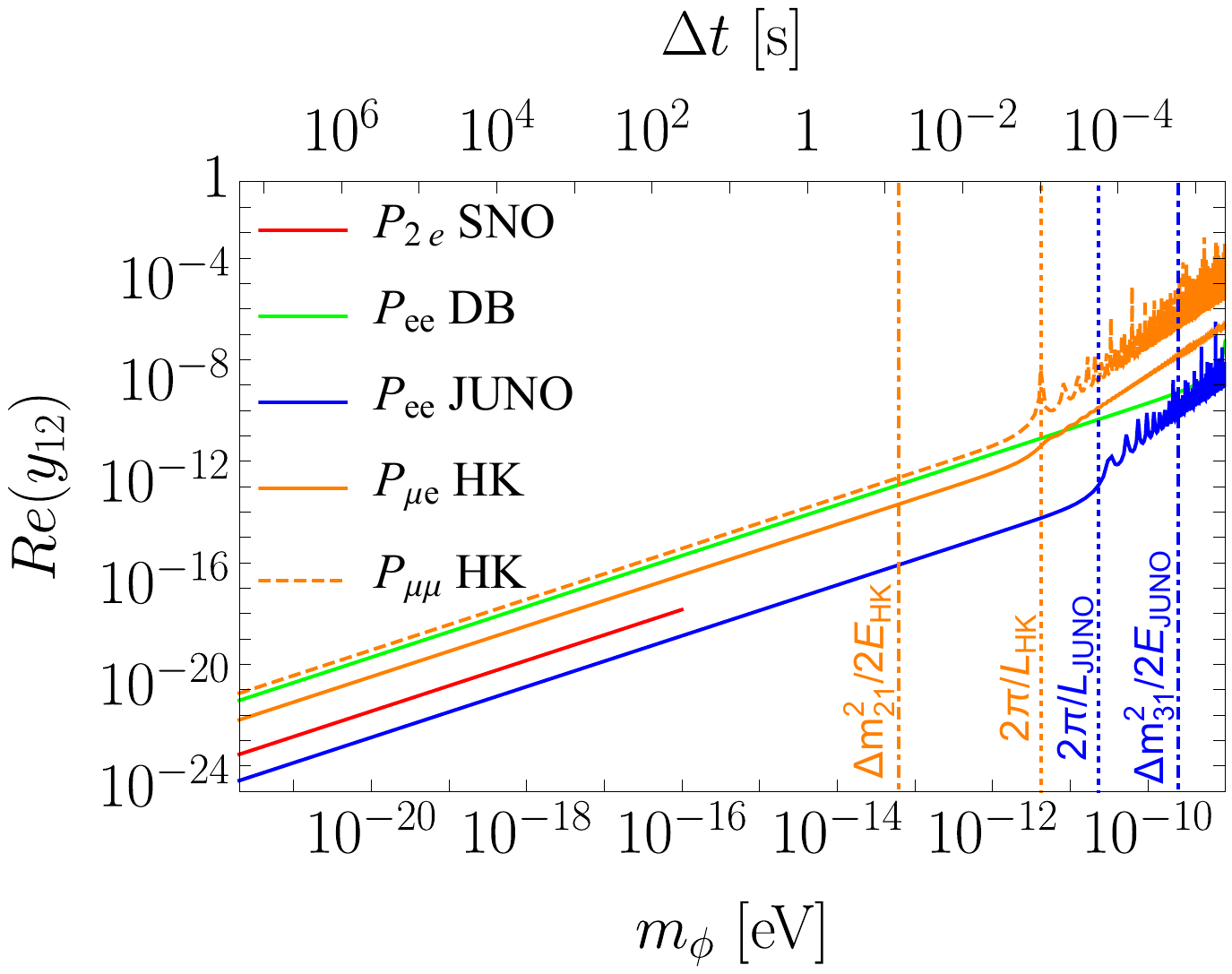}\includegraphics[scale=0.45]{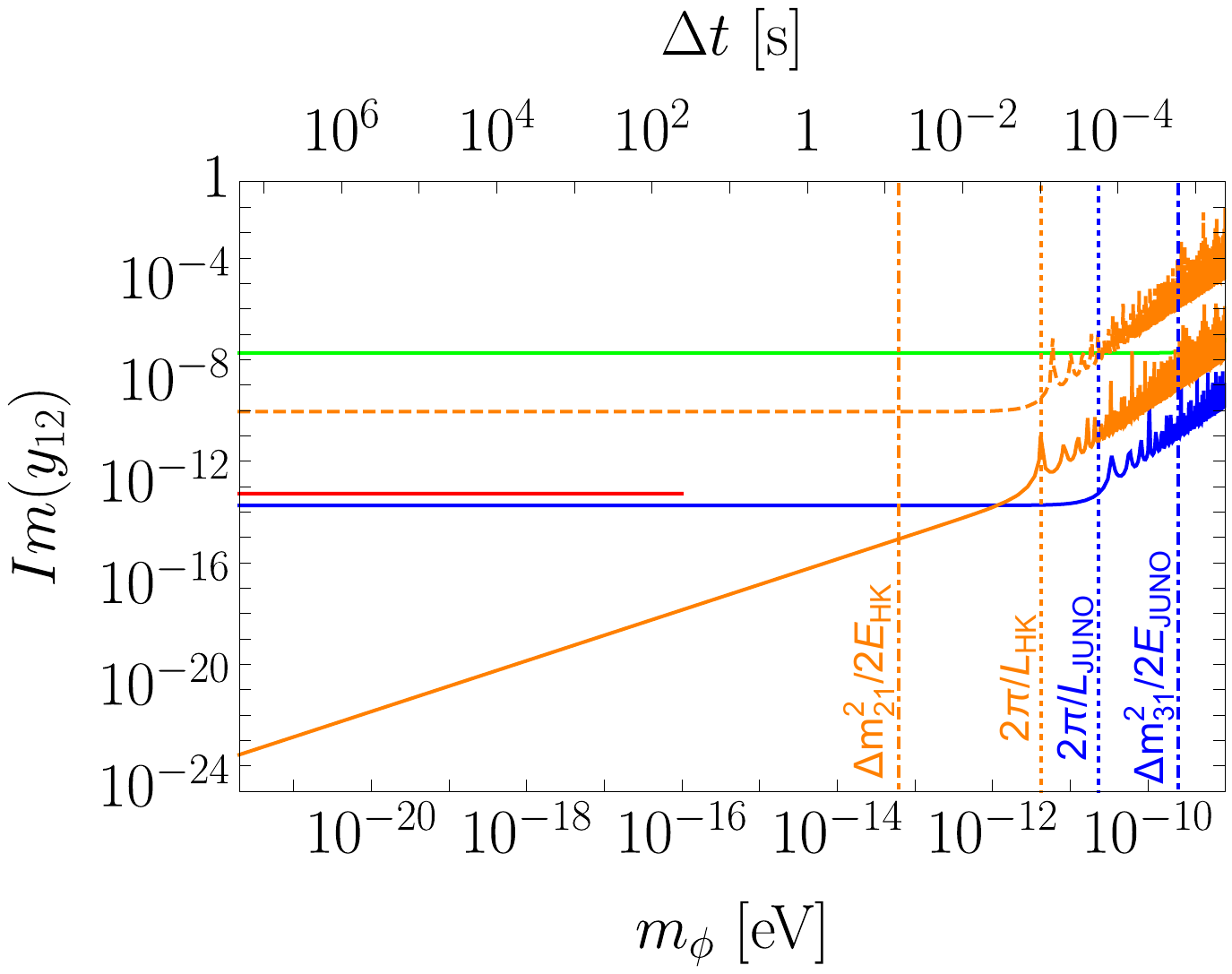}\\
		\includegraphics[scale=0.45]{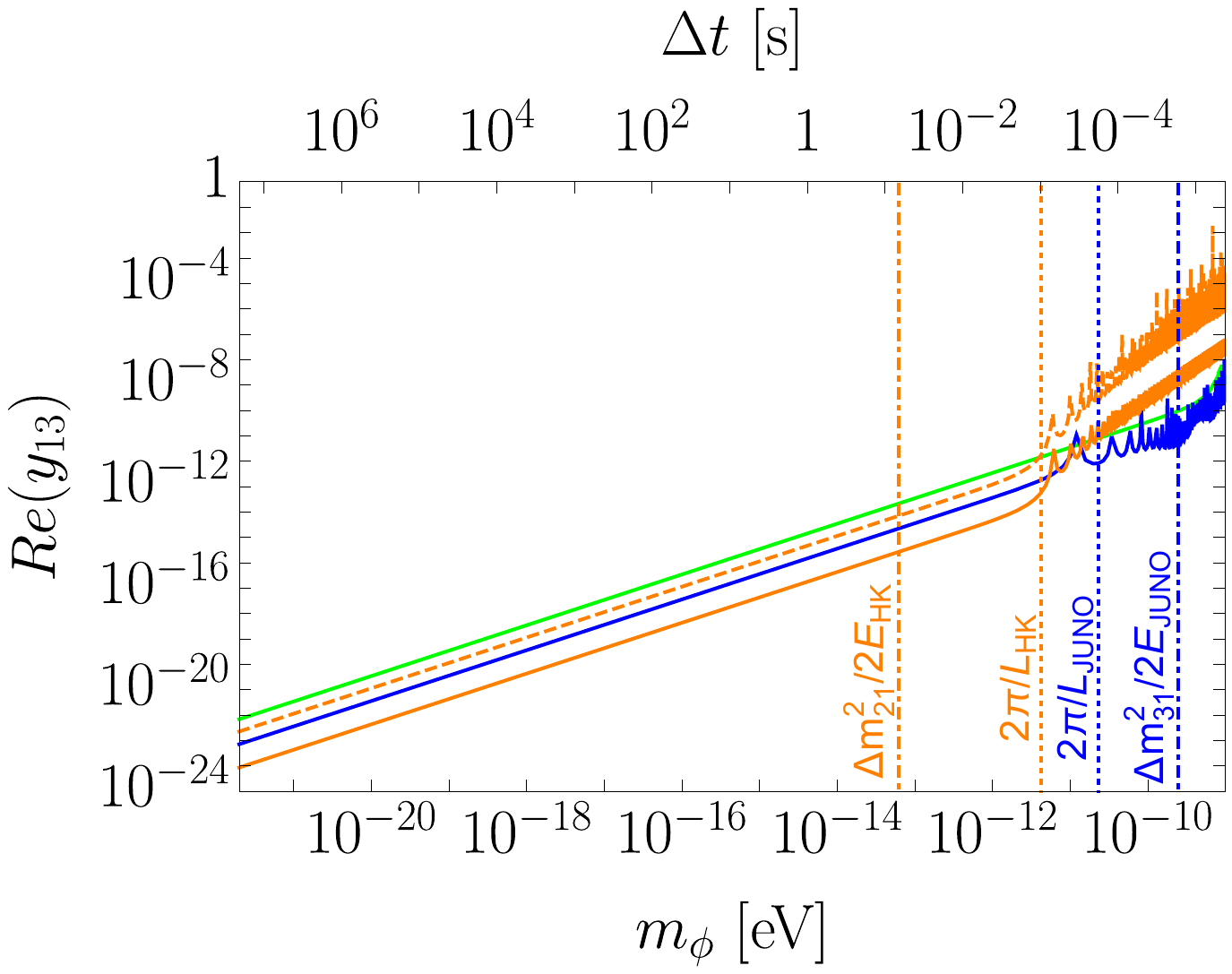}\includegraphics[scale=0.45]{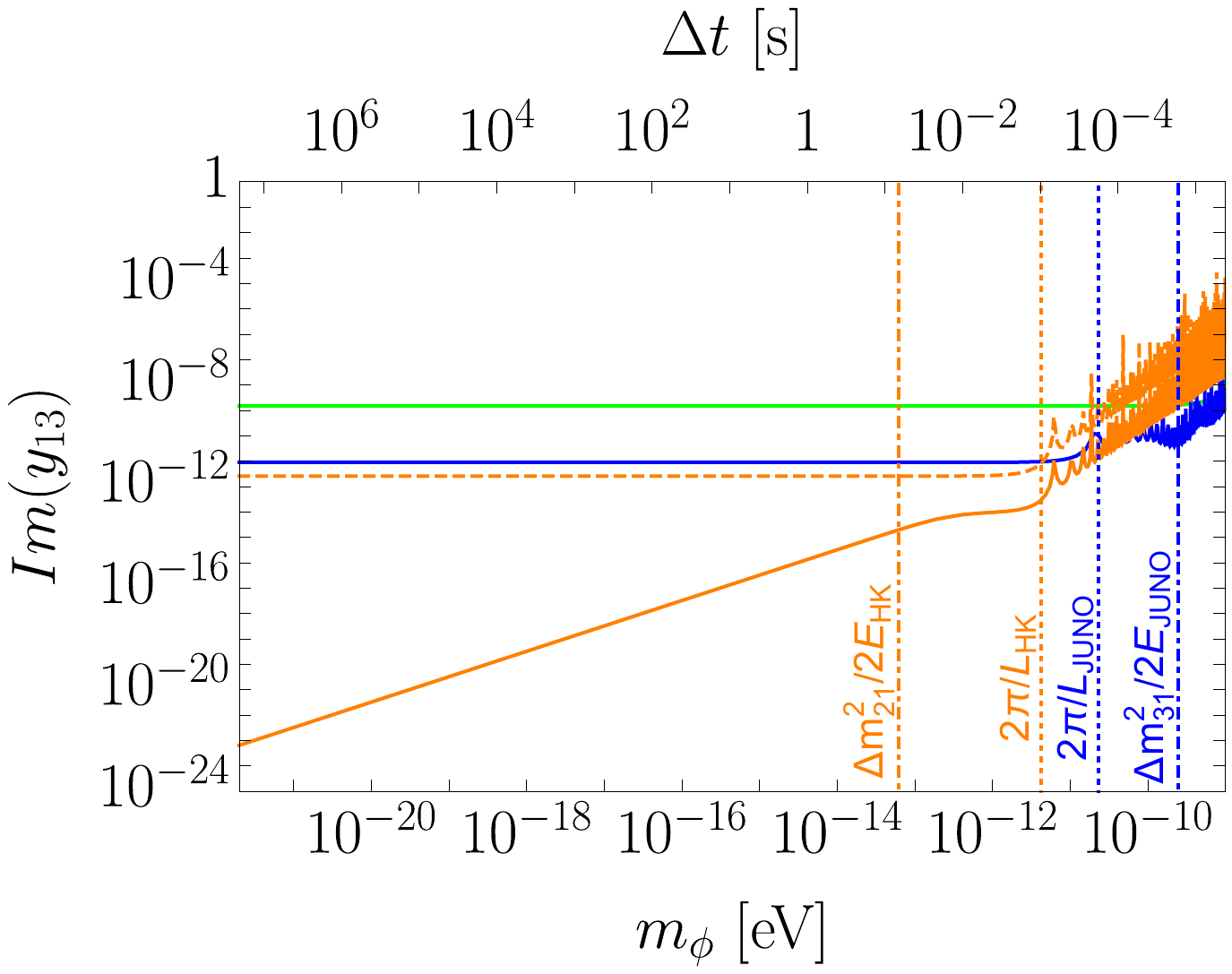}\\
		\includegraphics[scale=0.45]{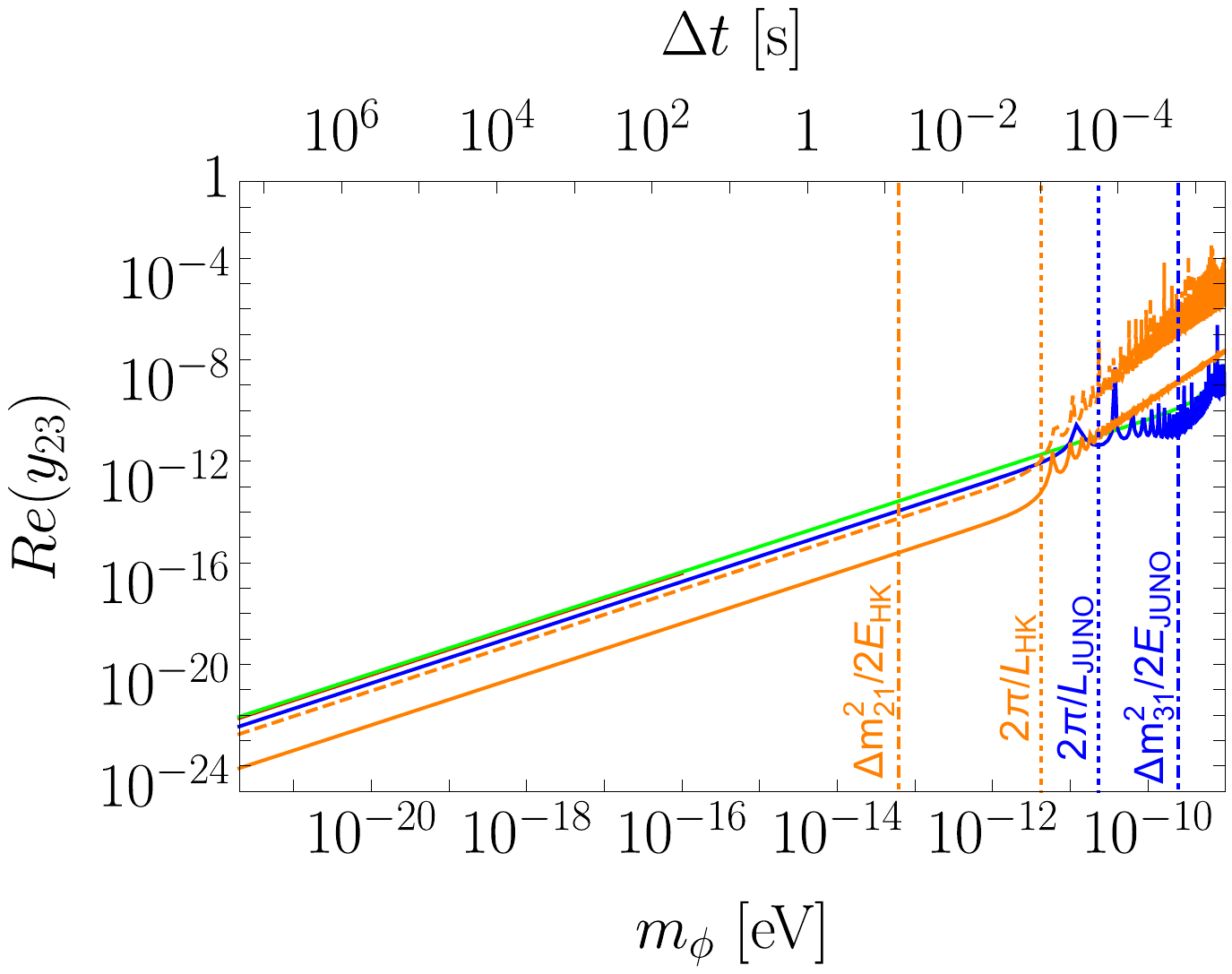}\includegraphics[scale=0.45]{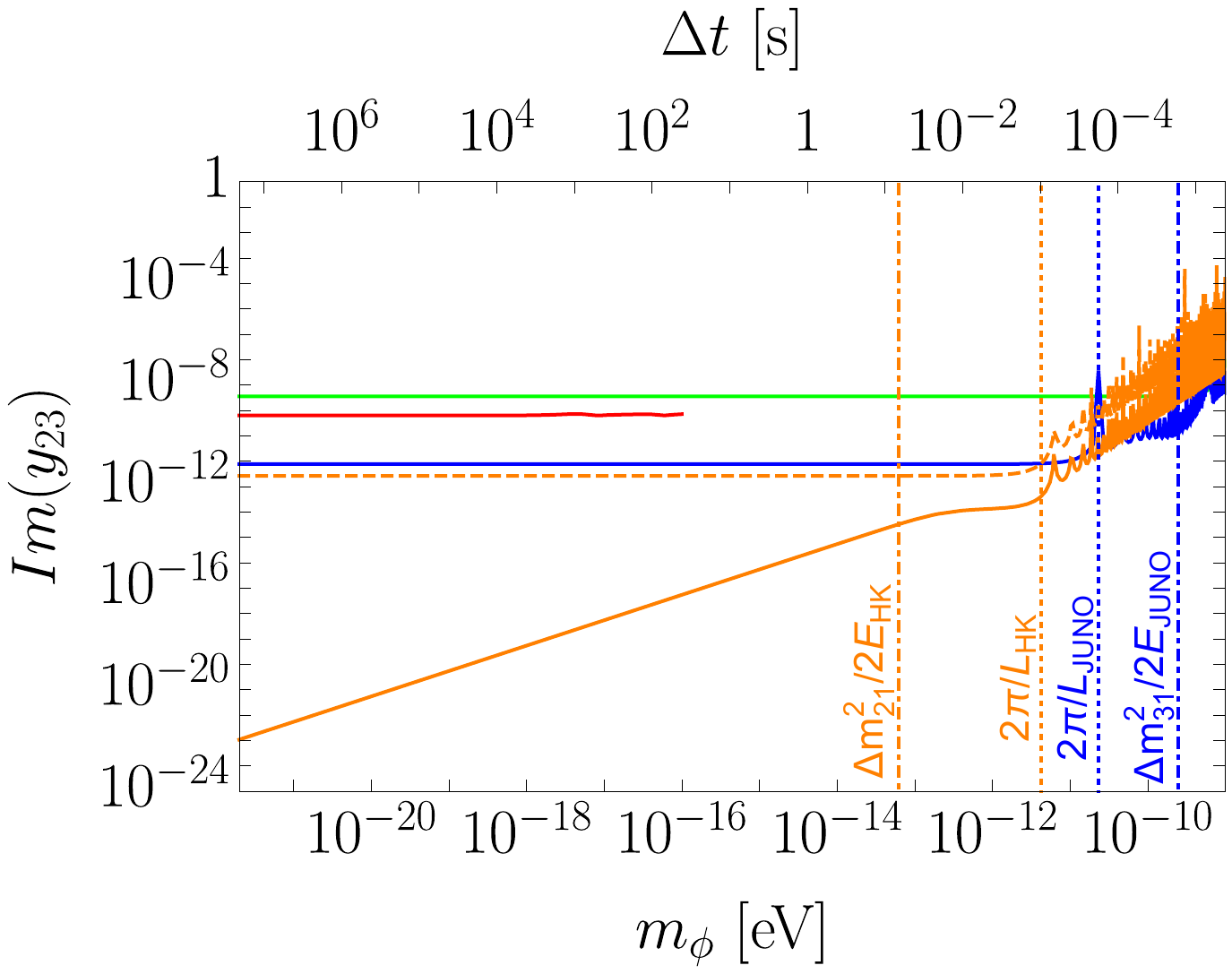}\\
		\includegraphics[scale=0.45]{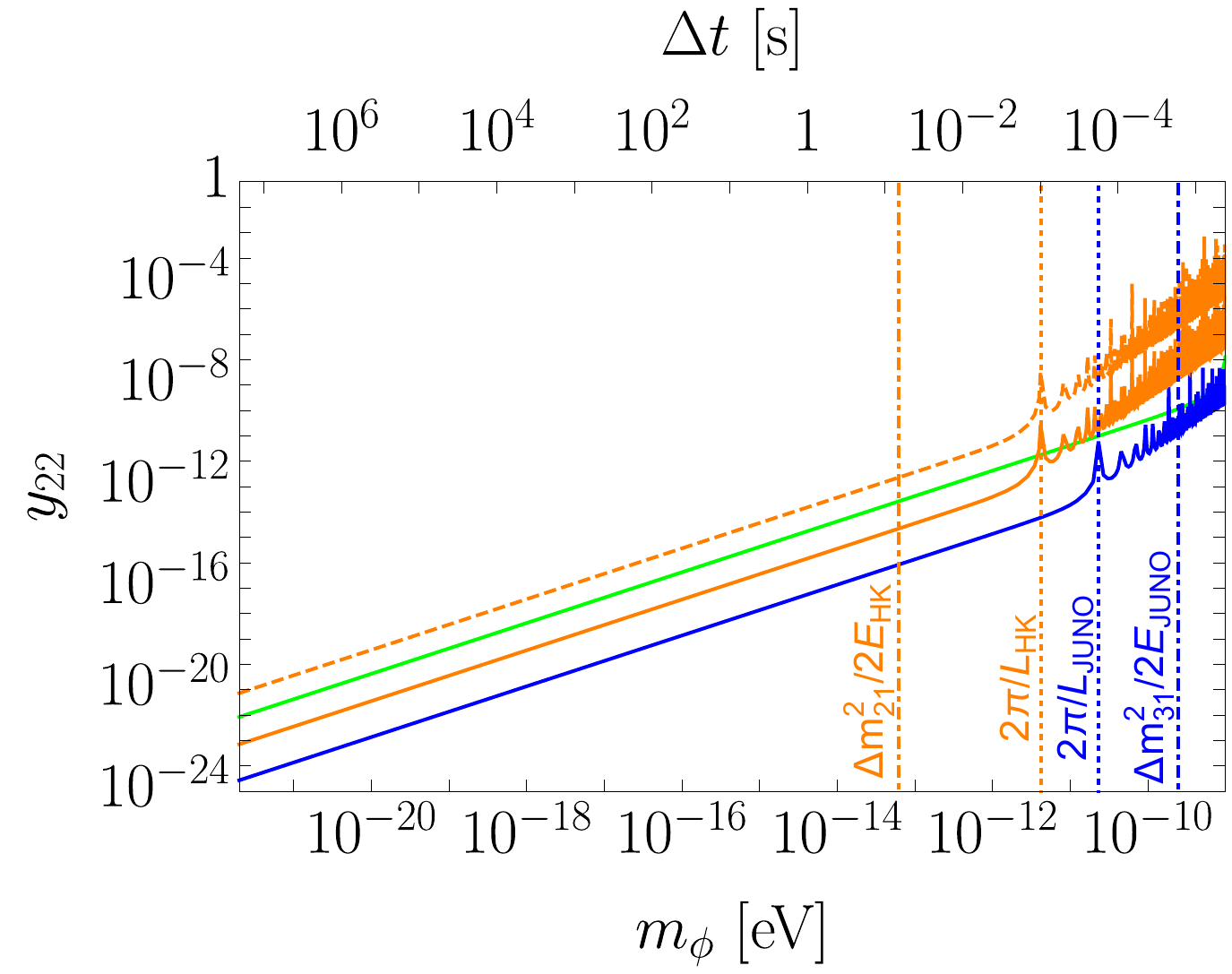}\includegraphics[scale=0.45]{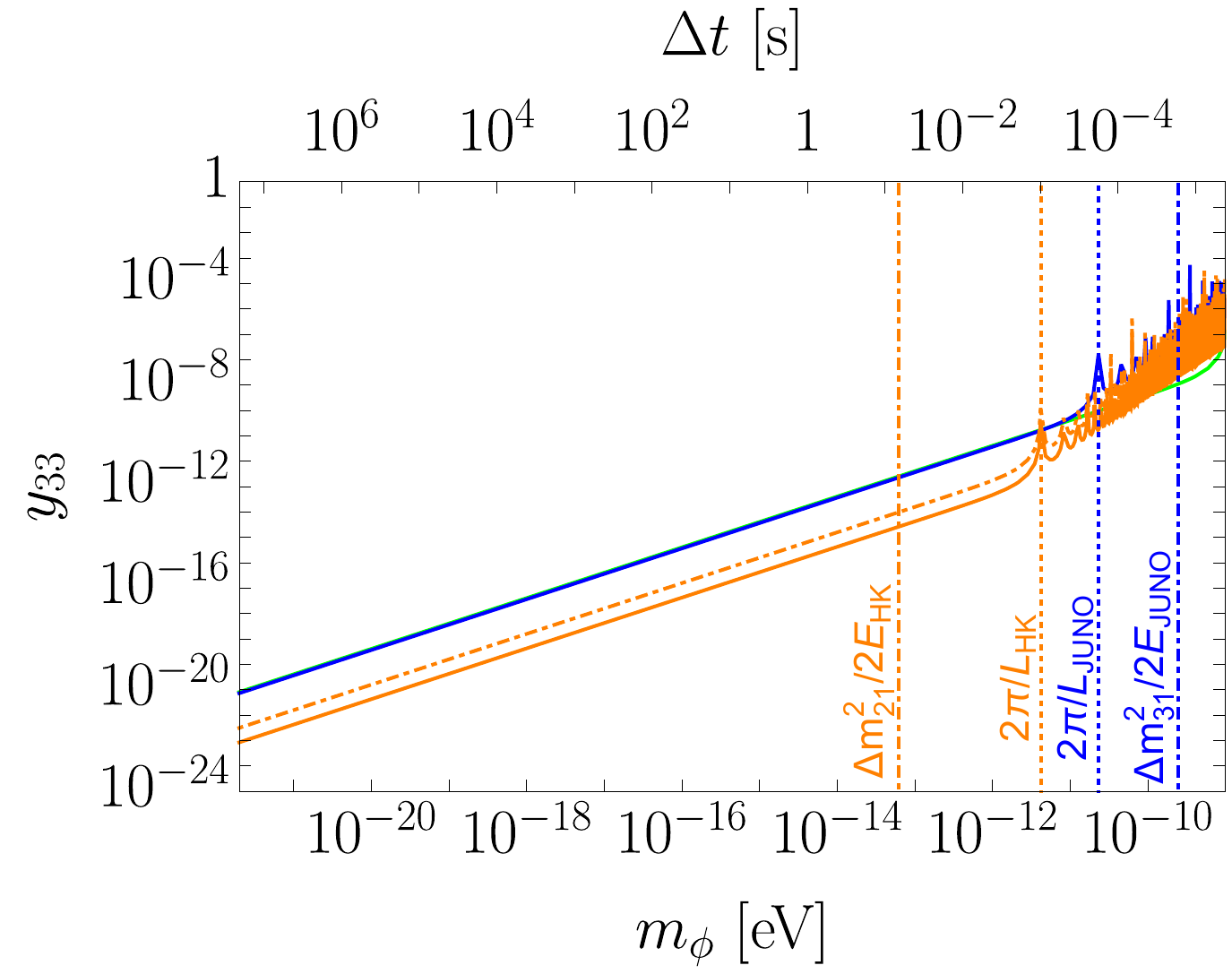}\\
		\par\end{centering}
	\caption{\label{fig: Experiments} Current bounds on the \ac{ULDM} parameter space from Daya-Bay and SNO experiments, alongside the projected sensitivities of JUNO and Hyper-K. To generate these plots we assumed normal hierarchy, $m_1=0$, and $\delta_{\text{CP}}=0$.}
\end{figure}

Accelerator neutrino experiments fire a beam of protons towards a target that produces charged pions. These can be manipulated into a narrow beam and decay mostly into $\nu_{\mu}$ or $\bar{\nu}_{\mu}$, depending if the positive or negative pions are focused. These experiments therefore measure $P_{\mu\mu}$ and $P_{\mu e}$ (or $P_{\bar{\mu}\bar{\mu}}$ and $P_{\bar{\mu}\bar{e}}$) simultaneously. We studied the DUNE, ESS (far and near detectors), and Hyper-K experiments, out of which the latter has produced overall slightly better sensitivities to the \ac{ULDM} parameters, and only these results are presented in Fig.~\ref{fig: Experiments}. Hyper-K is expected to collect approximately $2000$ to $4000$ signal events in each mode of neutrinos that travel $295$ km, over a run of 10 years~\cite{Hyper-KamiokandeWorkingGroup:2013hcb}. 

\subsection{Asymptotic behavior}
To qualitatively understand the various plots in Fig.~\ref{fig: Experiments}, let us consider the asymptotic behavior following from the analytical expressions we found in Sec.~\ref{sec:analytic_expressions}. Note that we only account for the first order corrections to the neutrino oscillation probability. Therefore, near the resonance $m_\phi=\Delta m^2/2E$~\cite{Losada:2022uvr} the result may be inaccurate, as higher order terms would yield considerable contributions.

Let us first discuss the behavior of the \ac{CPC} \ac{ULDM} effects (corresponding to the plots presenting the real parts of the couplings, assuming $\delta_{\rm{CP}}=0$). For couplings that are diagonal in the unperturbed mass basis, the correction to the transition or survival probability in Eq.~\eqref{eq:CPCdiagonal} is proportional to
\begin{align}
\Sigma P^{\left(1\right)} & \propto\sin\left(\frac{m_{\phi}L}{2}\right)\frac{2y_{ii}\phi_{0}}{Em_{\phi}}\propto\begin{cases}\frac{\text{Re}\left(y_{ii}\right)}{m_{\phi}} & m_{\phi}L\ll1\,,\\
\sin\left(\frac{m_{\phi}L}{2}\right)\frac{\text{Re}\left(y_{ii}\right)}{m_{\phi}^{2}} & m_{\phi}L\gg1\,.
\end{cases}
\end{align}
Therefore, in an experiment sensitive to a certain value of $P^{(1)}$, the corresponding diagonal coupling $y_{ii}$ that the experiment will be sensitive to is proportional to
\begin{align}
\text{Re}\left(y_{ii}\right) & \propto \begin{cases}
m_{\phi} & m_{\phi}L\ll1\,,\\
m_{\phi}^{2}/\sin\left(m_{\phi}L/2\right) & m_{\phi}L\gg1\,.
\end{cases}
\end{align}
We can see the transition between the two limits in the two bottom plots of Fig.~\ref{fig: Experiments}.

As expected, when the couplings are off-diagonal, the probability also depends on the neutrino oscillation frequency, and we obtain the following relation 
\begin{align}
\label{eq:sensRey}
\text{Re}\left(y_{ij}\right) & \propto\begin{cases}
m_{\phi} & L^{-1}\gg m_{\phi}\,,\\
m_{\phi}/\sqrt{A^2\cos^2\left(m_{\phi}L/2\right)+B^2\sin^2\left(m_{\phi}L/2\right)} & L^{-1}\ll m_{\phi}\ll\Delta m_{ij}^{2}/2E\,,\\
m_{\phi}^{2}/\sqrt{A^2\cos^2\left(m_{\phi}L/2\right)+B^2\sin^2\left(m_{\phi}L/2\right)} & L^{-1},\Delta m_{ij}^{2}/2E\ll m_{\phi}\,,
\end{cases} 
\end{align}
where $A$ and $B$ are determined by the coefficients of $\text{Re}(\kappa)$ and $\text{Im}(\kappa)$ in $P^{(1)}$, which depend on PMNS elements. If $A$ and $B$ are of similar magnitude, as is the case when $\text{Re}(\kappa)$ and $\text{Im}(\kappa)$ yield similar contributions to the probability, we expect the oscillations in $m_\phi$ to be suppressed. However, if one is much larger than the other, there are approximate poles of ``bad sensitivity" around the nodes of the bigger term, at which the bound follows the inverse of the smaller term (for example, if $A^2/B^2\gg1$, then around $m_\phi L/2 = n\pi$ the bound is proportional to $1/A$, while around $m_\phi L/2 = \pi/2+n\pi$ the bound is proportional to $1/B$). This is the case when either $\text{Re}(\kappa)$ or $\text{Im}(\kappa)$ dominate the probability. In the case of diagonal entries of $\hat{y}$, only $\text{Re}(\kappa)$ affects the probability, and therefore the bound oscillates rapidly for larger $m_\phi$. For off-diagonal $y$ entries, the modification to the \ac{CPC} survival probability is affected only by $\text{Re}(\kappa)$, while the transition probability is affected also by $\text{Im}(\kappa)$. Therefore, the sensitivity for off-diagonal \ac{CPC} couplings from measurements of survival probabilities will oscillate rapidly for larger $m_\phi$, while these oscillations are suppressed for transition probabilities.

The two extreme limits of Eq.~\eqref{eq:sensRey} are similar to the diagonal couplings case, while we may also obtain an in-between behavior if $\Delta m_{ij}^{2}L/\leri{2E}>1$. This region may be probed due to the neutrino mass hierarchy $\Delta m^2_{21}\ll \Delta m^2_{31},\Delta m^2_{32}$. The JUNO experiment is designed in a way such that $\Delta m^2_{21}L/\leri{2E}={\cal O}(1)$, and $\Delta m^2_{31}L/\leri{2E}\gg1$. This allows for the in-between region where $L^{-1}\ll m_\phi\ll\Delta m^2_{31}/\leri{4E}$. We can indeed see this region in the plots of $y_{13},y_{23}$, but not for $y_{12}$. 

The behavior of the \ac{CPV} couplings is similar to that of the \ac{CPC} couplings, with one exception: the 2-$\nu$ \ac{CPV} effect. The 2-$\nu$ \ac{CPV} is the only \ac{CPV} effect in survival probabilities, and its mass dependence follows that of $\text{Im}\leri{\kappa_{ij}}$ in Eq.~\eqref{eq:Imk}, and thus we expect that the bounds from survival probabilities follow
\begin{align}
\label{eq:sensImy}
\text{Im}\left(y_{ij}\right) & \propto\begin{cases}
\text{const}\leri{m_\phi} & L^{-1}\gg m_{\phi}\,,\\
m_{\phi}/\sqrt{A^2\cos^2\left(m_{\phi}L/2\right)+B^2\sin^2\left(m_{\phi}L/2\right)} & L^{-1}\ll m_{\phi}\ll\Delta m_{ij}^{2}/2E\,,\\
m_{\phi}^{2}/\sqrt{A^2\cos^2\left(m_{\phi}L/2\right)+B^2\sin^2\left(m_{\phi}L/2\right)} & L^{-1},\Delta m_{ij}^{2}/2E\ll m_{\phi}\,.
\end{cases} 
\end{align}
The main difference from the \ac{CPC} bounds is in the low mass region, where the bound flattens. This is a result of the fact that in the two generations case, as we discussed in Sec~\ref{sec:2_nu_CPV}, the only observable \ac{CPV} effect comes from the time-derivative of the \ac{ULDM}-induced phase. The 2-$\nu$ \ac{CPV} effect occurs also in transition probabilities, but overcomes the 2-$\nu$ \ac{CPV} effect in the low mass region if $\Delta m^2_{21}/E\ll m_\phi\ll\Delta m^2_{31}/E,\Delta m^2_{32}/E$ (this effectively sets $\Delta m^2_{21}=0$ which gives the two-neutrino picture). We can see this in-between behavior of the transition probability in Hyper-K for $y_{13}$ and $y_{23}$, but not for $y_{12}$. Since the 2-$\nu$ \ac{CPV} effect takes over at larger masses (for all off-diagonal couplings), the \ac{CPV} transition probabilities would inherit the behavior of the \ac{CPV} survival probabilities, which are solely affected by $\Im\leri{\kappa}$, and thus they would significantly oscillate as a function of $m_\phi$. 

\acresetall
\section{Conclusions}
\label{sec:conclusions}

The inclusion of a scalar \ac{ULDM} field that couples to neutrinos provides new rich phenomenology.
In this work, we have presented a generic analysis of the impact of time varying \ac{ULDM} interactions with neutrinos. For the case when the \ac{ULDM} rapidly oscillates during neutrino propagation, we have determined the corrections to the survival and transition probabilities for both \ac{CPC} and \ac{CPV} quantities. Some of the novel effects that appear are the following:
\begin{itemize}
\item A new \ac{CPV} effect is present even in the two neutrino-like case. We have also explicitly checked that in the three generation case this new effect is also present. Moreover, for three generations of neutrinos, the \ac{CPV} effect does not disappear in the limit of one of the $\Delta m_{ij}^{2} \rightarrow 0$, and both survival and transition probabilities are modified by the new \ac{CPV} effect.
\item The phenomena we are interested in presents oscillatory behavior and thus we can look at the spectral power of the neutrino probabilities for angular frequencies defined by $m_{\phi}$ to determine the parameters in the model.
\end{itemize}
Interestingly, we have identified that several various neutrino oscillation experiments can be sensitive to the fast variations of the \ac{ULDM} field. Given this we have determined the bounds on \ac{CPC} and \ac{CPV} couplings from the neutrino survival probability measurements at Daya Bay and SNO. Furthermore, we have determined the sensitivity to measure the couplings at future reactor (JUNO) and long baseline neutrino oscillation experiments (DUNE, ESS and Hyper-K).

\section*{Acknowledgements}
YN is the Amos de-Shalit chair of theoretical physics, and is supported by grants from the Israel Science Foundation (grant number 1124/20), the United States-Israel Binational Science Foundation (BSF), Jerusalem, Israel (grant number 2018257), by the Minerva Foundation (with funding from the Federal Ministry for Education and Research), and by the Yeda-Sela (YeS) Center for Basic Research. The work of GP is supported by grants from BSF-NSF, Friedrich Wilhelm Bessel research award, GIF, ISF, Minerva, SABRA - Yeda-Sela - WRC Program, the Estate of Emile Mimran, and The Maurice and Vivienne Wohl Endowment. IS~is supported by a fellowship from the Ariane de Rothschild Women Doctoral Program.

\newpage
\appendix

\section{Two Generations}
\label{appendix:2nuP}
For diagonal \ac{ULDM} couplings, $\hat{y}_{ii}$, and $k\neq i$, the survival probability is modified by
\begin{align}
P^{(1)}_{\alpha\alpha}\leri{t}& =-\sin\leri{m_\phi \leri{t-\frac{L}{2}-t_0}}\sin\leri{\frac{m_\phi L}{2}}\frac{\sin^2{2\theta}\hat{y}_{ii} m_i\phi_0}{ E m_\phi}\sin{\leri{\frac{\Delta m^{2}_{ik}L}{2E}}}\,,
\label{eq:2nuPii}
\end{align}
where $\theta$ is the mixing angle between the unperturbed mass eigenstates and interaction eigenstates.
For non-diagonal \ac{ULDM} couplings $\hat{y}_{ij}$, with $i\neq j$
\begin{align}
P^{(1)}_{\alpha\alpha}& =\frac{\phi_0 \sin{4\theta} }{E\leri{\leri{\frac{\Delta m^{2}_{ij}}{2E}}^2-m_\phi^2}}\times\nonumber\\
\times\Big[&\text{Im}\leri{\hat{y}_{12}}\leri{m_1-m_2}\cos\leri{m_\phi \leri{\leri{t-L-t_0}+\frac{L}{2}}}\leri{\frac{\Delta m^{2}_{21}}{4E}\sin{\frac{\Delta m^{2}_{21}L}{2E}}\sin{\frac{m_\phi L}{2}}-m_\phi \sin^2{\frac{\Delta m^{2}_{21}L}{4 E}}\cos{\frac{m_\phi L}{2}}}\nonumber\\
+&\text{Re}\leri{\hat{y}_{12}}\leri{m_1+m_2}\sin\leri{m_\phi \leri{\leri{t-L-t_0}+\frac{L}{2}}}\leri{ \frac{m_\phi}{2}\sin{\frac{\Delta m^{2}_{21}L}{2E}}\sin{\frac{m_\phi L}{2}}-\frac{\Delta m^{2}_{21}}{2E}\sin^2{\frac{\Delta m^{2}_{21}L}{4 E}}\cos{\frac{m_\phi L}{2}}}\Big ]\,.\label{eq:2nuPij}
\end{align}

\section{Mean value of Rayleigh power at the modulation frequency}\label{appendix: Rayleigh}
Recall the time scales of the problem:
\begin{enumerate}
	\item $\tau_{e}$: The running time of the experiment.
	\item $\tau_{s}=\frac{N_{\nu}}{\tau_{e}}$: The average spacing between
	events, where $N_{\nu}$ is the total number of measured events.
	\item $\tau_{r}$: The resolution of the clock that times the events.
\end{enumerate}
Therefore, during the experiment running time, there are $N_t=\frac{\tau_e}{\tau_r}$ clock ticks, each with small duration $\tau_r$. At each clock tick we define the function:
\begin{equation}
h\left(t_{i}\right)=\left\{ \begin{matrix}1 & \text{event detected,}\\
0 & \text{no event detected.}
\end{matrix}\right.
\end{equation}
We calculate the expectation value of the Rayleigh power spectrum, namely
\begin{align}
\left\langle z\left(f\right)\right\rangle  & =\frac{2}{N_\nu}\leri{\left\langle \left|\sum_{i=1}^{N_t}h\left(t_{i}\right)\cos\left(2\pi t_{i}f\right)\right|^{2}\right\rangle +\left\langle \left|\sum_{i=1}^{N_t}h\left(t_{i}\right)\sin\left(2\pi t_{i}f\right)\right|^{2}\right\rangle}, 
\end{align}
were $f$ is the modulation frequency of the probability modulation, corresponding to $m_\phi/2\pi$. We calculate the expectation value of a generic function $F\left(h\left(t_{i}\right),t_{i}\right)$ in the sense that
\begin{equation}
\left\langle F\left(h\left(t_{i}\right),t_{i}\right)\right\rangle =F\left(1,t_{i}\right)\cdot P\left(h\left(t_{i}\right)=1\right)+F\left(0,t_{i}\right)\cdot P\left(h\left(t_{i}\right)=0\right),
\end{equation}
where the probability that an event is detected is given by
\begin{align}
P\left(h\left(t_{i}\right)=1\right) & =\frac{N_{\nu}}{N_t}\left[1+\epsilon\sin\left(2\pi t_i f\right)\right]=N_{\nu}\frac{\tau_{\text{r}}}{\tau_{\text{e}}}\left[1+\epsilon\sin\left(2\pi t_i f\right)\right],
\end{align}
where $\epsilon$ is the amplitude of the modulation. We assumed for simplicity that the phase of the modulation is zero, such that it appears as a sine, but the final result holds for a generic phase. Let us start with calculating the expectation value of the sine term. Notice that
\begin{equation}
\label{eq: var, epsilon}\left\langle \left|\sum_{i=1}^{N_t}h\left(t_{i}\right)\sin\left(2\pi t_{i}f\right)\right|^{2}\right\rangle =\text{Var}\left(\sum_{i=1}^{N_t}h\left(t_{i}\right)\sin\left(2\pi t_{i}f\right)\right)+\left\langle \left|\sum_{i=1}^{N_t}h\left(t_{i}\right)\sin\left(2\pi t_{i}f\right)\right|\right\rangle ^{2}\,.
\end{equation}
Calculating the variance term first, 
\begin{align}
\text{Var}\left[h\left(t_{i}\right)\sin\left(2\pi t_{i}f\right)\right] & =\left\langle h^{2}\left(t_{i}\right)\sin^{2}\left(2\pi t_{i}f\right)\right\rangle -\left\langle h\left(t_{i}\right)\sin\left(2\pi t_{i}f\right)\right\rangle ^{2}=\leri{P\leri{t_i}-P^2\leri{t_i}}\sin^{2}\left(2\pi t_{i}f\right)\nonumber\\
 &\approx\frac{N_{\nu}}{N_t}\left[1+\epsilon\sin\left(2\pi t_{i}f\right)\right]\sin^{2}\left(2\pi t_{i}f\right)
)\,.\end{align}
In the last approximation we assume $\frac{N_{\nu}}{N_t}\ll1$, which is equivalent to the statement that the resolution of the clock is much better than the typical time between events. The variance of sum, is simply the sum of variances, since in our case the measurements at different times are uncorrelated, thus
\begin{align}
\text{Var}\left(\sum_{i=1}^{N_t}h\left(t_{i}\right)\sin\left(2\pi t_{i}f\right)\right) & =\sum_{i=1}^{N_t}\frac{N_{\nu}}{N}\left[1+\epsilon\sin\left(2\pi t_{i}f\right)\right]\sin^{2}\left(2\pi t_{i}f\right)\,.
\end{align}
Changing variables:
\begin{align}
t_{n} & \rightarrow n\tau_{\text{res}}\,,\\
0 & <n\leq N_t\,,
\end{align}
we obtain
\begin{align}
\text{Var}\left(\sum_{i=1}^{N_t}h\left(t_{i}\right)\sin\left(2\pi t_{i}f\right)\right) & =\frac{N_{\nu}}{N_t}\left[\sum_{n=1}^{N_t}\sin^{2}\left(2\pi n\tau_{\text{res}}f\right)+\epsilon\sum_{n=1}^{N_t}\sin^{3}\left(2\pi n\tau_{\text{res}}f\right)\right]\,.
\end{align}
The first term yields
\begin{align}
\sum_{n=1}^{N_t}\sin^{2}\left(2\pi n\tau_{\text{res}}f\right) & \approx\int\limits _{0}^{N_t}dz\sin^{2}\left(2\pi z\tau_{\text{res}}f\right)
 =\frac{N_t}{2}\left[1-\text{sinc}\left(4\pi N_t\tau_{\text{res}}f\right)\right]
 \approx\frac{N_t}{2}\,,
\end{align}
where we use the approximation
\begin{equation}
N_t\tau_{\text{res}}f=\tau_{\exp}f\gg1.~\label{eq:approxf}
\end{equation}
The second term gives
\begin{align}
\sum_{n=1}^{N_t}\sin^{3}\left(2\pi n\tau_{\text{res}}f\right) & \approx\int\limits _{0}^{N_t}dz\sin^{3}\left(2\pi z\tau_{\text{res}}f\right) =\frac{2N_t\left(2+\cos\left(2fN_t\pi\tau_{\text{res}}\right)\right)\sin^{3}\left(fN_t\pi\tau_{\text{res}}\right)}{3}\text{sinc\ensuremath{\left(fN_t\pi\tau_{\text{res}}\right)}} \,,
\end{align}
and may therefore be neglected under the approximation in Eq.~\eqref{eq:approxf}.\\
The second term in Eq. \eqref{eq: var, epsilon} is
\begin{align}
\left\langle \left|\sum_{i=1}^{N_t}h\left(t_{i}\right)\sin\left(2\pi t_{i}f\right)\right|\right\rangle ^{2} & =\left|\sum_{i=1}^{N_t} P(t_i)\sin\left(2\pi t_{i}f\right)\right|^2=\left[\sum_{i=1}^{N_t}\frac{N_{\nu}}{N_t}\left[1+\epsilon\sin\left(2\pi t_{i}f\right)\right]\sin\left(2\pi t_{i}f\right)\right]^{2} =\frac{\epsilon^{2}}{4}N_{\nu}^{2}.
\end{align}
Finally we get
\begin{align}
\left\langle \left|\sum_{i=1}^{N_t}h\left(t_{i}\right)\sin\left(2\pi t_{i}f\right)\right|^{2}\right\rangle  & =\frac{N_{\nu}}{2}+\frac{\epsilon^{2}}{4}N_{\nu}^{2}\,,\\
\left\langle \left|\sum_{i=1}^{N_t}h\left(t_{i}\right)\cos\left(2\pi t_{i}f\right)\right|^{2}\right\rangle  & =\frac{N_{\nu}}{2}\,,
\end{align}
and thus
\begin{align}
\left\langle z\left(f\right)\right\rangle  & =2+\frac{\epsilon^{2}}{2}N_{\nu}\,.
\end{align}

\bibliographystyle{JHEP}
\bibliography{2-nu_CPV}

\end{document}